\documentclass[%
 reprint,
superscriptaddress,
 amsmath,amssymb,
 aps,
]{revtex4-2}

\usepackage{graphicx}% Include figure files
\usepackage{dcolumn}% Align table columns on decimal point
\usepackage{bm}% bold math
\usepackage{color}
\usepackage{amsmath}
\newcommand{\Kn}{\mathrm{Kn}}
\begin{document}

\preprint{APS/123-QED}

\title{Data-Driven Constitutive Relation Reveals Scaling Law for Hydrodynamic Transport Coefficients}
\thanks{https://doi.org/10.1103/PhysRevE.107.015104}% Force line breaks with \\

\author{Candi Zheng}%
 \email{Corresponding author:  czhengac@connect.ust.hk}
\affiliation{%
Department of Mechanics and Aerospace Engineering, Southern University of Science and Technology, Xueyuan Rd 1088, Shenzhen, China 
}%
\affiliation{%
  Department of Mathematics,
  Hong Kong University of Science and Technology, Clear Water Bay, Hong Kong SAR, China
}%
 \author{Yang Wang}%
 \email{yangwang@ust.hk}
 \affiliation{%
  Department of Mathematics,
  Hong Kong University of Science and Technology, Clear Water Bay, Hong Kong SAR, China
}%
  \author{Shiyi Chen}%
 \email{ chensy@sustech.edu.cn}
\affiliation{%
Department of Mechanics and Aerospace Engineering, Southern University of Science and Technology, Xueyuan Rd 1088, Shenzhen, China 
}%
\date{\today}% It is always \today, today,

\begin{abstract}
Finding extended hydrodynamics equations valid from the dense gas region to the rarefied gas region remains a great challenge. The key to success is to obtain accurate constitutive relations for stress and heat flux. Data-driven models offer a new phenomenological approach to learning constitutive relations from data. Such models enable complex constitutive relations that extend Newton's law of viscosity and Fourier's law of heat conduction by regression on higher derivatives. However, the choices of derivatives in these models are ad-hoc without a clear physical explanation. We investigated data-driven models theoretically on a linear system. We argue that these models are equivalent to non-linear length scale scaling laws of transport coefficients. The equivalence to scaling laws justified the physical plausibility and revealed the limitation of data-driven models. Our argument also points out that modeling the scaling law could avoid practical difficulties in data-driven models like derivative estimation and variable selection on noisy data. We further proposed a constitutive relation model based on scaling law and tested it on the calculation of Rayleigh scattering spectra. The result shows our data-driven model has a clear advantage over the Chapman-Enskog expansion and moment methods.

\end{abstract}

\maketitle

%\tableofcontents

\section{Introduction} Multiscale physics is widely encountered in fluid dynamics \cite{Chen1998LATTICEBM}, soft matter systems \cite{Peter2009MultiscaleSO}, and quantum chemistry \cite{Lin2006QMMMWH}. One of the typical multiscale physics problems is rarefied gas dynamics \cite{Kogan1969RarefiedGD}. Rarefied gas flow simulation is known to be difficult due to the non-negligible dynamics at the mesoscopic scale. Simulation resolving these scales is computationally expensive for continuous and transitional flows, such as the Direct Simulation Monte Carlo (DSMC) \cite{Bird1994} method. Instead, extended hydrodynamics equations at a coarse-grained macroscopic scale are efficient substitutes to reduce the computational cost. What lies within the heart of extended hydrodynamics is constitutive relations. Constitutive relations summarize mesoscopic scale dynamics as macroscopic phenomena, such as viscosity and heat conduction. Traditionally they are modeled by perturbation or polynomial expansion around the equilibrium of dense gas. The perturbation models, such as the Burnett-type equations \cite{agarwal2001beyond, GarcaColn2008BeyondTN}, utilize high-order spatial derivatives according to the Hilbert-Chapman-Enskog expansion \cite{Hilbert1912, Chapman1916,enskog1917kinetische}. However, difficulties exist in stability \cite{bobylev1982chapman} and non-guaranteed convergence \cite{McLennan1965ConvergenceOT}. The polynomial expansion methods are represented by the Grad moment method \cite{Grad1949} and its extensions \cite{Struchtrup2011} modifying the equilibrium distribution with orthogonal polynomials. Nevertheless, issues appear in unphysical solutions \cite{weiss1995continuous} and hyperbolicity \cite{Torrilhon2009HyperbolicME}.
The traditional methods' perturbation or expansion nature limited their viability near dense equilibrium, constraining their applicable Knudsen numbers range \cite{Struchtrup2011}.

Data-Driven models offer a new phenomenological approach to obtaining machine-learned constitutive relations from data. Compared to perturbation or expansion from existing theory, data provids an alternative source of information and is expected to expand the applicable range of extended hydrodynamics equations \cite{Hana2019}. There have been attempts to learn constitutive relations from mesoscopic results \cite{Zhang2020} or to find proper moment equations \cite{Hana2019}. Data-Driven models are also used in related areas such as learning the unknown governing of physical systems \cite{Raissi2017, Rudy2017, Schaeffer2017, Raissi2019}, simulating physical dynamics \cite{Long2018, Long2019, lecun1995convolutional} and solving the Boltzmann equation \cite{DalSanto2020}. These attempts have proven the concept of data-driven modeling. However, the advantage over traditional models like Chapman-Enskog and Grad moment method hasn't been established yet. Limitations for data-driven models include derivative estimation \cite{baydin2018automatic}, determining input quantities (variable selection) \cite{Schaeffer2017}, and modeling across a range of Knudsen numbers. Besides, the rather ad-hoc linear or neural network regression in data-driven models lack a clear physical explanation.

In this paper, we seek the physical explanation of data-driven models by investigating linear systems. We focus on the conservation laws and analyze data-driven constitutive relation models that extend Newton's viscosity law and Fourier's heat conduction law. We argue that these linear models are equivalent to non-linear length scale scaling laws of viscosity and heat conduction coefficients. These length scale scaling laws describe the change of viscosity and heat conduction coefficients, as we concern with dynamics at different length scales described by Knudsen numbers. The equivalence between data-driven constitutive relations and scaling laws justified the physical plausibility of data-driven models.

Based on our argument, we suggest modeling scaling laws explicitly in data-driven models. In doing so, we could involve high-order derivatives implicitly in constitutive relations without calculating them. It helps to avoid practical difficulties in data-driven models like derivative estimation and variable selection. We further modeled the constitutive relation based on our suggestion. 

We apply our model to calculate the Rayleigh scattering spectra as the numerical benchmark. The Rayleigh scattering have been well studied \cite{Yip1964} and used in Lidar wind measurement \cite{Fiocco1968}. However, it remains difficult to correctly model the spectra shape in the transition region for today's extended-hydrodynamic equations \cite{Wu2020}. The numerical results show that our data-driven model can capture the spectra shape at the transition region. To our knowledge, it is the first time that the data-driven hydrodynamic model significantly outperforms the traditional Chapman-Enskog expansion and Grad moment methods.

\section{Methods}

We consider the linearized extended hydrodynamics for one-dimensional homogeneous rarefied ideal gas. The hydrodynamics equations govern the dynamics of gas. The most important hydrodynamics equations are mass, momentum, and energy conservation laws. They form a one-dimensional (1D) linear system of density $\rho$, velocity $v$, and temperature $T$, respectively. The non-dimensionalized linear system for conservation laws is
 \begin{equation}\label{Non-dim Macroscopic Equations}
	\begin{split}
		\frac{\partial  {\rho}}{\partial  {t}} + \frac{\partial  {v}}{\partial  {x}}     &= 0\\
		\frac{\partial  {v}}{\partial  {t}} + \frac{\partial  {T}}{\partial  {x}} + \frac{\partial  {\rho}}{\partial  {x}} &=   -\mbox{Kn}  \frac{\partial   {\sigma}}{\partial  {x}}  \\
		\frac{3 }{2}\frac{\partial  {T}}{\partial  {t}} + \frac{\partial  {\rho}}{\partial  {x}} &= -  \frac{15 }{4 } \mbox{Kn}\frac{\partial   {q}}{\partial  {x}},
	\end{split}
\end{equation} 
in which $\mbox{Kn}$ is the {\em Knudsen number} describing how rarefied the gas is. Detailed descriptions of non-dimensionalization and the definition of the Knudsen number are in Appendix \ref{Ap A}. 

However, the hydrodynamics equations are not closed with two extra unknown terms: the stress ${\sigma}$ and the heat flux ${q}$ that encodes the mesoscopic dynamics. To close the equations, constitutive relations that model the stress and the heat flux with known quantities are necessary.

\subsection{Data-Driven Constitutive Relations and Its Equivalence with Scaling Laws}

We adopt a general form of data-driven constitutive relations consisting of derivatives of various orders similar to other data-driven models for physical systems \cite{Raissi2017, Rudy2017, Schaeffer2017, Raissi2019, Long2018, Long2019}. It is also motivated by the Hilbert-Chapman-Enskog expansion. The Hilbert-Chapman-Enskog expansion is a systematic way to generate constitutive relations for conservation laws at small Knudsen numbers. The leading order of expansion yields the well-known Newton's law of viscosity and Fourier's law of heat conduction and defines the viscosity coefficient $\mu_0$ and heat conduction coefficient $\kappa_0$. However, they are not valid for rarefied gas effects at large Knudsen numbers \cite{chapman1990mathematical}. For large Knudsen numbers, higher-order expansions extend the capability of constitutive relations by incorporating high-order spatial derivatives of density, velocity, and temperature. If we consider linear systems, these high-order spatial derivatives are combined linearly by coefficients determined by the Hilbert-Chapman-Enskog expansion. However, the Hilbert-Chapman-Enskog expansion guarantee neither convergence nor stability of the system \cite{agarwal2001beyond}. Similar to the Hilbert-Chapman-Enskog expansion, we consider constitutive relations linear combinations of high-order spatial derivatives. However, we aim to determine combinations coefficients via a data-driven regression approach. Therefore we adopt the following general form of the data-driven constitutive relation \begin{equation} \label{The Linear Model on Derivatives: stress_and_flux general}
	\begin{split}
		 {\sigma} &= - \sum_{n=1}^{\infty}  \left( {a}_n \frac{\partial^n  {v}}{\partial  {x}^n}  + {c}_n \frac{\partial^n  {\rho}}{\partial  {x}^n} +{e}_n \frac{\partial^n  {T}}{\partial  {x}^n}\right) \\
		 {q} &=-\sum_{n=1}^{\infty} \left(  {b}_n\frac{\partial^n  {T}}{\partial  {x}^n}+{d}_n \frac{\partial^n  {\rho}}{\partial  {x}^n}+{f}_n \frac{\partial^n  {v}}{\partial  {x}^n} \right),
	\end{split}
\end{equation}
where $x$ is the non-dimensional spatial coordinate,  $a_n,b_n,c_n,d_n,e_n,f_n$ are unknown {\em regression coefficients}. The constitutive relation \eqref{The Linear Model on Derivatives: stress_and_flux general} has the same functional form obtained from Hilbert-Chapman-Enskog expansion since both are combinations of high-order spatial derivatives. But the {\em regression coefficients} in \eqref{The Linear Model on Derivatives: stress_and_flux general} are to be determined via a data-driven approach.

There are practical difficulties in directly applying constitutive relation \eqref{The Linear Model on Derivatives: stress_and_flux general}. First is the problem in variable selection. This problem arises because we only have limited data in practice to determine the infinitely many regression coefficients in \eqref{The Linear Model on Derivatives: stress_and_flux general}. Consequently, we could only determine a selected subset of regression coefficients. Choosing the best subset of regression coefficients is a challenging variable selection problem we wish to avoid. The second problem is density estimation. Constitutive relation \eqref{The Linear Model on Derivatives: stress_and_flux general} contains high-order spatial derivatives, which are difficult to estimate in practice. A naive attempt at estimating high-order spatial derivatives using the finite difference method requires a highly dense mesh and is very sensitive to noise. It completely fails on data generated by the DSMC method since they contain strong statistical noise. Finally, constitutive relation \eqref{The Linear Model on Derivatives: stress_and_flux general} does not guarantee the stability of hydrodynamic equations. The reason is there are no constraints on entropy production yet to respect the second law of thermodynamics. Fortunately, it turns out that reformulating the problem in the Fourier space with proper constraints on entropy production enables us to bypass the practical difficulties in variable selection and derivative estimation. 

%Fortunately, proper constraints on the entropy production reveal the inherent equivalent between constitutive relation \eqref{The Linear Model on Derivatives: stress_and_flux general} and scaling laws of transport coefficients, which enable us to bypass the practical difficulties in variable selection and derivative estimation. 

Now we reformulate the constitutive relations with the help of the Fourier transform and entropy production constraints. Fourier transform allows us to convert the derivatives in constitutive relations into algebraic expressions. Meanwhile, constraints on entropy production eliminate undesired terms and imaginary parts that appear in the Fourier transformation. The outline of the reformulation is as follows: Firstly, the entropy constraint reduces the constitution relations to the form that stress $\sigma$ consists of only velocity derivatives and heat flux $q$ consists of only temperature derivatives. This is because the stress and velocity, the same as heat flux and temperature, must be correlated to produce non-increasing entropy as follows.
\begin{equation} \label{Hydrodynamic Fluctuations:Total entropy change rate, expand 1}
\begin{split}
	\dot{s} & = -\frac{q}{T^2}\partial_x T - \frac{\sigma}{T} \partial_x v  ,
\end{split}
\end{equation}
where $\dot{s}$ is the entropy change rate per volume \cite{Landau1980hf}. The only possibility is stress depends on velocity only, the same as heat flux depending on temperature, since density, velocity, and temperature are statistically independent \cite{Landau1980ipt}. Secondly, the non-increasing constraint on the entropy production eliminates undesired imaginary parts in the Fourier transform of constitutive relations. This constraint requires that each Fourier mode of the density, velocity, and temperature must produce non-negative entropy. It is necessary if we wish the linear system to be stable. As a result, constitutive relations are expressed as a summation of infinite polynomial series in the Fourier space. Finally, collecting and reformulating the summation in the constitutive relations leads to the following constitutive relations in the Fourier space
\begin{equation} \label{The Linear Model on Derivatives: stress_and_flux, not flux, Fourier}
	\begin{split}
		  \tilde{\sigma}( k) & = -i k \frac{4}{3} \frac{\mu(k)}{ \mu_0} \tilde{v}( k); \quad \mu( k) \ge 0 \\
		 \tilde{q}(  k) &=-i k \frac{\kappa(k)}{\kappa_0} \tilde{T}( k); \quad \kappa(k) \ge 0 ,
	\end{split}
\end{equation}
where $\mu_0$ is the viscosity coefficient, $\kappa_0$ is the heat conduction coefficient, $k$ is the non-dimensional wavenumber for each Fourier mode, $\tilde{\sigma}( k),\tilde{q}(  k), \tilde{v}( k), \tilde{T}( k)$ are corresponding spatial Fourier transforms of $\sigma, q, v, T$. The detailed derivation from the constitutive relation \eqref{The Linear Model on Derivatives: stress_and_flux general} to \eqref{The Linear Model on Derivatives: stress_and_flux, not flux, Fourier} are shown in Appdendix \ref{Ap: B}. The derivation also shows that the functions $\mu(k), \kappa(k)$ are even functions satisfy the natural constraints $\mu_0 = \lim_{k\rightarrow 0} \mu(k) $ and $\kappa_0 =\lim_{k\rightarrow 0} \kappa(k)$. As we will discuss later, they describe the length scaling law for viscosity and heat conduction coefficient. Therefore we have shown that the data-driven constitutive relation \eqref{The Linear Model on Derivatives: stress_and_flux general} transformed into the form \eqref{The Linear Model on Derivatives: stress_and_flux, not flux, Fourier} containing scaling laws under the constraint of non-increasing entropy. This established the equivalence between data-driven constitutive relations and scaling laws.

The functions $\mu(k), \kappa(k)$ are the length scaling laws of viscosity and heat conduction coefficients.  They describe the relative change of viscosity and heat conduction coefficients w.r.t length scale changes of the system. This is because $k$ is closely related to the {\em Knudsen number} $\mbox{Kn} = l/L$ that characterize the length scale of a rarefied gas system, in which $l$ is the mean free path of gas molecules and $L$ is the representative length scale of the system. Particularly, as defined in Appendix \ref{Ap A}, the Knudsen number of a Fourier mode is proportional to its non-dimensionalized wavenumber $k$
\begin{equation} \label{KnK}
 \Kn \propto |k|.
\end{equation}
Therefore the even functions $\mu(k), \kappa(k)$ are also functions of the Knudsen number, hence are length scaling laws. These scaling laws could be measured experimentally \cite{Michalis2010}. However, we can not use such experimental results directly because the definition of the Knudsen number is not unified but varies according to the experiment setting. Alternatively, scaling laws could be learned through a data-driven approach from data like fluctuation spectra \cite{GhaemMaghami1980RayleighBrillouinSO} containing information on viscosity and heat conduction.

Scaling laws $\mu(k), \kappa(k)$ in \eqref{The Linear Model on Derivatives: stress_and_flux, not flux, Fourier} are much easier to be determined than regression coefficients in \eqref{The Linear Model on Derivatives: stress_and_flux general}. These coefficients may lead to divergence at large Knudsen numbers, making it ill-conditioned to determine regression coefficients valid for large Knudsen numbers. Instead, we could learn scaling laws $\mu(k), \kappa(k)$ uniformly from data at various Knudsen numbers without worrying about convergence. Learning scaling laws also eliminates the demand in variable selection, which refers to choosing a subset of regression coefficients. It is because all regression coefficients are now summarized in the function $\mu(k), \kappa(k)$. Moreover, learning scaling laws is robust against noisy data since it avoids using estimated derivatives in constitutive relations \eqref{The Linear Model on Derivatives: stress_and_flux general}. Therefore learning the scaling laws, compared to regression coefficients in \eqref{The Linear Model on Derivatives: stress_and_flux general}, avoid practical difficulties in convergence issue, variable selection, and derivative estimation.

\subsection{Modeling Scaling Laws using Neural Network}

One difficulty that remains is learning scaling laws from data turns out to be a non-convex optimization problem that is difficult to solve. We overcome it by  approximating scaling laws $\mu(k), \kappa(k)$ using neural networks, taking advantage of their stochastic optimization technique designed for non-convex optimizations \cite{Goodfellow2015DeepL}. 

Neural network modeling functions $\mu, \kappa$ must be constrained to obtain correct asymptotics and symmetry for hydrodynamics. Asymptotically, the function values of $\mu, \kappa$ must be specified to the equilibrium values $\mu_0, \kappa_0$ at $\Kn=0$ to guarantee the constitutive relation's consistency with the Navier-Stokes equation. In addition, we couple $\kappa$ and $\mu$ together 
\begin{equation} \label{Fix Pr}
\kappa(k) = \frac{5 k_B}{2m} \frac{\mu(k)}{\mbox{Pr}},\quad \mbox{Pr} = \frac{2}{3}  
\end{equation}
to constraint the Prandtl number $\mbox{Pr}$ to the Chapman-Enskog result $\mbox{Pr}=\frac{2}{3}$. While this coupling is not necessary, we find it accelerates the learning process without undermining the accuracy in practice. As for symmetry, homogeneity in space also requires the scaling laws to be even functions of $k$. Homogeneity means there is no preferred direction in space. Hence the direction in space coordinate or the corresponding wavenumber $k$ should not make a difference in the scaling laws. For the one-dimensional case, the direction of k is its sign. Therefore, the scaling laws must be even functions independent of the sign of $k$.

To satisfy all these constraints, we design the following non-dimensional constrained neural network for $\mu$ satisfying $M(\Kn) = \frac{4}{3} \frac{\mu(\Kn)}{ \mu_0}$ with the architecture
\begin{equation} \label{The neural network}
\begin{split}
 M(\Kn)  &= \frac{4}{3} \left(1 +   \mathbf{W}_2 \cdot \mbox{Tanh} (\mathbf{W}_1 \cdot \mathcal{H}(\mbox{20Kn}) )\right) \\
\mathcal{H}(x)&= \frac{1}{2} x^2, x < 1;\quad |x|-\frac{1}{2},x \ge 1 , 
\end{split}
\end{equation}
in which $\mbox{Kn}$ are proportional to $|k|$ as in \eqref{KnK}, $\mathbf{W}_{1}, \mathbf{W}_{2}$ are the one-dimension weight vectors of the neural network, with the activation function $\mbox{Tanh}$ acting element-wise on the vector input. The function $M(\mbox{Kn})$ is even and satisfies $M(0) = \frac{4}{3}$ and $M'(0) = 0$. It guarantee the consistency with the NS equation. With the modeling of the scaling laws, we are prepared to investigate the capability of scaling laws in describing rarefied gas dynamics.

\begin{figure}
\includegraphics[width=3.5in]{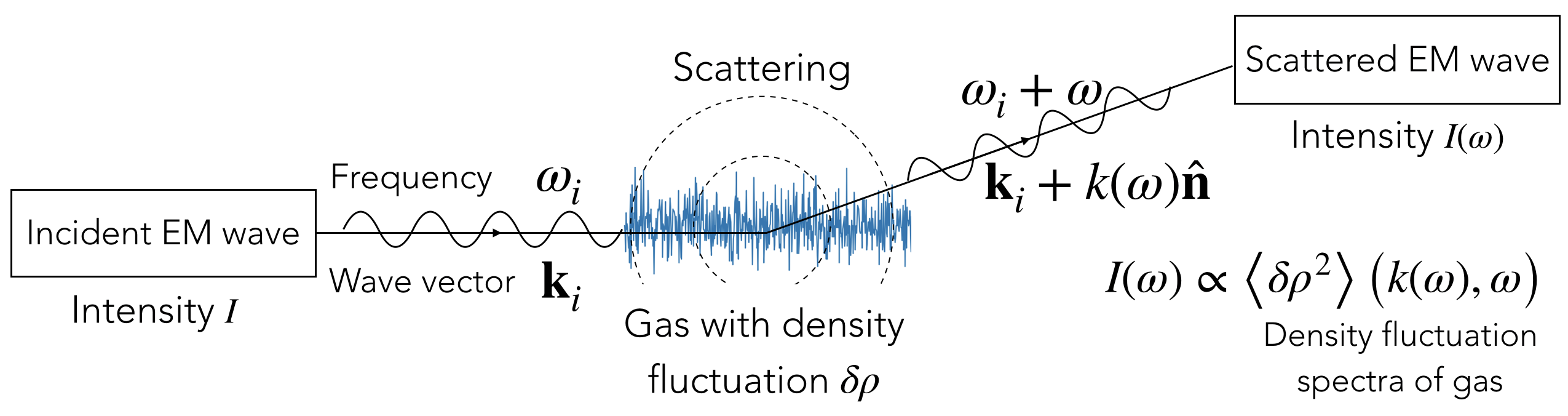}% Here is how to import EPS art
\caption{\label{fig:Rayleigh} Rayleigh scattering of electromagnetic(EM) waves with wave vector $\bold{k}_i$ and frequency $\omega_i$ through gas with density fluctuation $\delta \rho$. $\hat{\bold{n}}$ is a unit vector. The intensity $I(\omega)$ of scattered EM wave with frequency shift $\omega$ is the Rayleigh scattering spectra. The Rayleigh scattering spectra are proportional to and determined by the density fluctuation spectra $\left<\delta\rho^2\right>$. Therefore to compute the Rayleigh scattering spectra, we need only to compute the density fluctuation spectra of gas.}
\end{figure}

\subsection{Rayleigh Scattering as Benchmark case}

We will test the capability of scaling laws $\mu, \kappa$ in describing rarefied gas dynamics by calculating the Rayleigh scattering spectra. The {\em Rayleigh scattering} describes the refraction of electromagnetic waves (EM waves) passing through media with stochastic density fluctuation \cite{Pecora1964, jackson1999classical, landau2013electrodynamics}. Such fluctuation usually appears as density fluctuation waves and happens spontaneously with the thermal motion of gas molecules. The {\em Rayleigh scattering spectra} are defined as the intensities $I(\omega)$ of scattered EM waves after the Rayleigh scattering with frequency shifts $\omega$, as shown in Fig \ref{fig:Rayleigh}. They are proportional to the density fluctuation spectra of gas
\begin{equation} \label{Rayleigh Scattering Spectra: spectral density for with space-time correlation function, omega4}	
	I(\omega) \propto  \left<\delta \rho^2\right>(k(\omega), \omega), 
\end{equation}
where $k$ is the wavenumber change of the scattered EM wave determined observation position and incident wave frequency, $\left<\delta\rho^2\right>$ is the {\em density fluctuation spectra}, which describes the intensity of density fluctuation waves at each wavenumber $k$ and frequency $\omega$. A detailed description on the relation between the Rayleigh scattering spectra and density fluctuation spectra is shown in Appendix \ref{Ap: C}. As a consequence of the proportionality between the Rayleigh scattering spectra and density fluctuation spectra, calculating the Rayleigh scattering spectra only needs to compute the density fluctuation spectra of the gas media. 
 
\begin{figure}
\includegraphics[width=3.5in]{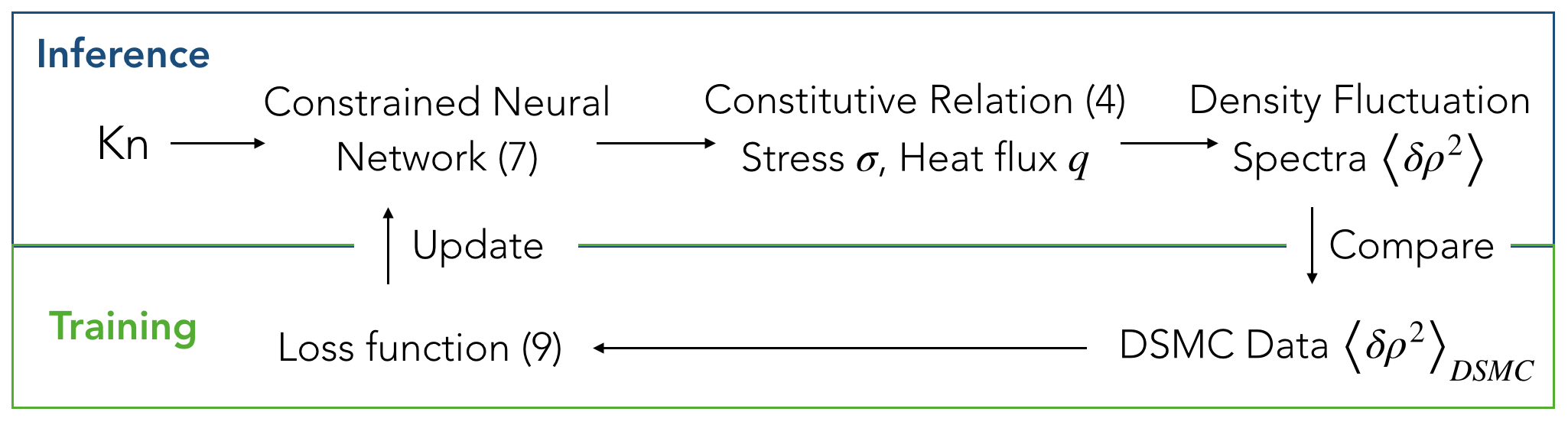}% Here is how to import EPS art
\caption{\label{fig:Flowchart} The flow chart of our model to calculate the density fluctuation spectra $\left<{\delta \rho}^2\right>$.}
\end{figure}

The density fluctuation spectra $\left<{\delta \rho}^2\right>(k, \omega)$ describe the amplitude of density fluctuation waves caused by the collective motions of gas molecules. The wavenumber $k$ in the density fluctuation spectra specifies the wavelength of density fluctuation waves. It also sets the Knudsen number of density fluctuation waves since the wavenumber $k$ is proportional to the Knudsen number. Given a Knudsen number by specifying $k$,  the spectra $\left<{\delta \rho}^2\right>(k, \omega)$ could be calculated from macroscopic governing equations \eqref{Non-dim Macroscopic Equations} using constitutive relation \eqref{The Linear Model on Derivatives: stress_and_flux, not flux, Fourier} (details in Appendix \ref{Ap: D}). Hence the values of scaling laws $\mu, \kappa$ in the constitutive relation affect the shape of spectra $\left<{\delta \rho}^2\right>$ as a function of $\omega$. It means that the density fluctuation spectra contain information on scaling laws which we aim to extract by training the neural network models. In practice, we train the neural network modeling scaling laws on density fluctuation spectra data $\left<\delta\rho^2\right>_{dsmc}$ computed by the DSMC method (Appendix \ref{Ap: E}).

The density fluctuation spectra are not enough to confirm the capability of scaling laws in describing rarefied gas dynamics. It is because there is the risk of overfitting. Overfitting refers to the neural network learning the scaling law by rote from density fluctuation spectra. In other words, the neural network learns a scaling law that fails in predicting quantities other than density fluctuation spectra. To eliminate the risk of overfitting, we need to prepare test data to examine the neural network's generalization ability: the ability to predict quantities that the neural network has not seen in the training process.

We examine the generalization ability of the neural network on test data consisting of velocity fluctuation spectra $\left<{ v}^2\right>(k, \omega)$. Similar to density fluctuation spectra, velocity fluctuation spectra describe the amplitude of velocity fluctuation waves caused by the collective motions of gas molecules. Velocity fluctuation spectra serve as ideal test data because of the following reasons: first, velocity fluctuation is consistent with the scaling law discussed in our paper since velocity fluctuation also obeys the hydrodynamic equations \eqref{Non-dim Macroscopic Equations}; second, velocity fluctuation corresponds to a different physical scenario compared to density fluctuation. In detail, velocity fluctuations are solved from the hydrodynamic equations with an initial condition ( \eqref{AP_ini delta condition v} in Appendix \ref{Ap: D}  ) completely different from density fluctuation ( \eqref{AP_ini delta condition} in Appendix \ref{Ap: D} ). The `consistent but different' characteristic of velocity fluctuations makes them ideal for examing the generalization ability of our neural-network-modeled scaling laws.

\subsection{Training the Neural Network on Density Fluctuation Spectra Data}

We train the neural network models for scaling laws on the density fluctuation spectra data $\left<{\delta \rho}^2\right>$. Specifically, it refers to learning the weight vectors $\mathbf{W}_{1}, \mathbf{W}_{2}$ in the neural network \eqref{The neural network} from data. This requires a loss function as the learning target. In our case, the loss function compares the difference between the observed spectra $\left<\delta\rho^2\right>_{dsmc}$ and predicted spectra $\left<\delta\rho^2\right>$. The former are training data obtained from the DSMC computation (Appendix \ref{Ap: E}), while the latter are the predictions of the governing equation. We define the loss function for any input weight vector $\mathbf{W} = \mathbf{W}_{1}, \mathbf{W}_{2}$ as
\begin{equation} \label{The loss function}
	\begin{split}
    L(\mathbf{W})& =\mathbb{E}_{\Kn \sim U} \mathbb{E}_{\omega \sim p(\omega|\Kn)} \\ 
    & \left| \left<\delta\rho^2\right>_{dsmc} (\Kn,\omega) -\left<\delta\rho^2\right> (\Kn,\omega;\mathbf{W})) \right|^2  ,
	\end{split}
\end{equation}
in which the predicted spectra $\left<\delta \rho^2\right>$ is a function on the weight vectors $\mathbf{W}$ because it depends on the neural network $M(\Kn)$. The symbol $\mathbb{E}_{\Kn \sim U}$ represents taking the expectation numerically by sampling $\Kn$ from a uniform distribution $U$. Meanwhile, $ \mathbb{E}_{\omega \sim p(\omega|\Kn)}$ represents taking the expectation by sampling $\omega$ from a conditional distribution $p(\omega|k)$, which is proportional to the amplitude of the DSMC spectra. Sampling $\omega$ in this way makes the sample point lies more in the peak region. After defining the loss function, we use the ADAM \cite{Kingma2015} optimizer to minimize the loss function and determine the weight vectors (Appendix \ref{AP: F}).

We take extra caution on the finite domain effect of the DSMC computed spectra data. The DSMC simulates gas confined in 1-D space of finite domain length $L_d$. However, we aim to compute the density fluctuation spectra for Fourier modes with infinite spatial span. Therefore, finite domain length inevitably affects the spectra, especially for Fourier modes with a wavelength comparable to domain length. Such finite domain effect is proportional to the mean free path and domain length ratio $\frac{l}{L_d}$ which vanishes as $L_d$ tends to infinity. As a solution, we use a large domain length much greater ($>200$ times) than the mean free path of the gas, eliminating the finite domain effect in the DSMC computed spectra.

In total, the numerical experimental setting could be divided into two processes: inference and training. The inference process calculates the density fluctuation spectra using the governing equations with the constitutive relations  \eqref{The Linear Model on Derivatives: stress_and_flux, not flux, Fourier}. The constitutive relations contain neural networks $M, K$ defined in \eqref{Fix Pr} and \eqref{The neural network} with weights to be determined. The training process determines the weights of neural networks by minimizing the loss function  \eqref{The loss function}. The flow chart Fig. \ref{fig:Flowchart} summarizes the entire procedure.
\begin{figure}
\includegraphics[width=3.5in]{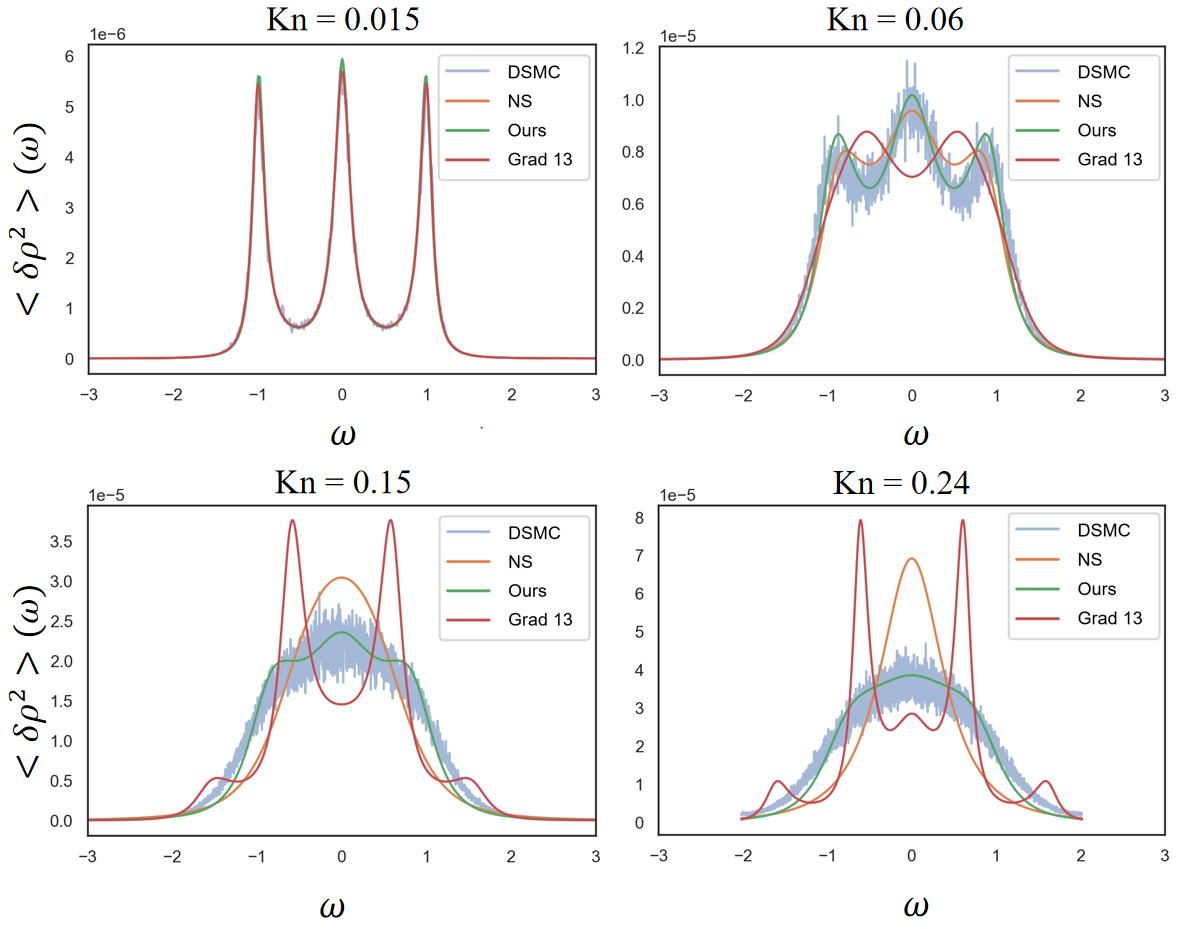}% Here is how to import EPS art
\caption{\label{spectra}Comparison between spectra calculated using DSMC, the NS equation, the Grad 13 method, and our model for various Knudsen numbers. At a small Knudsen number, The spectra consist of three peaks corresponds to entropy fluctuation and pressure fluctuation. As the Knudsen number increase, these peaks disappear gradually and blur into a bell shape. The result showed that our model calculated spectra match the DSMC result much better than the NS equation and Grad 13 method, especially in the high Knudsen number region.}
\end{figure}

\section{Results} 

We compare the density fluctuation spectra calculated by our model with the results of the NS equation and the Grad 13 method. For various Knudsen numbers, spectra $\left<\delta \rho^2\right>(\tilde{\omega})$ are shown in Fig. \ref{spectra} as function of the non-dimensionalized frequency $\tilde{\omega}$. At a small Knudsen number, all models give consistent spectra. However, at large Knudsen number, our model result matches accurately with the DSMC result, while the shape and amplitude of the NS equation and Grad 13 moments method deviate.  Therefore, compared with the NS equation and the Grad 13 method, our model gives the most accurate spectra which are close to the DSMC result in both shape and amplitude.

\begin{figure}
\includegraphics[width=3.3in]{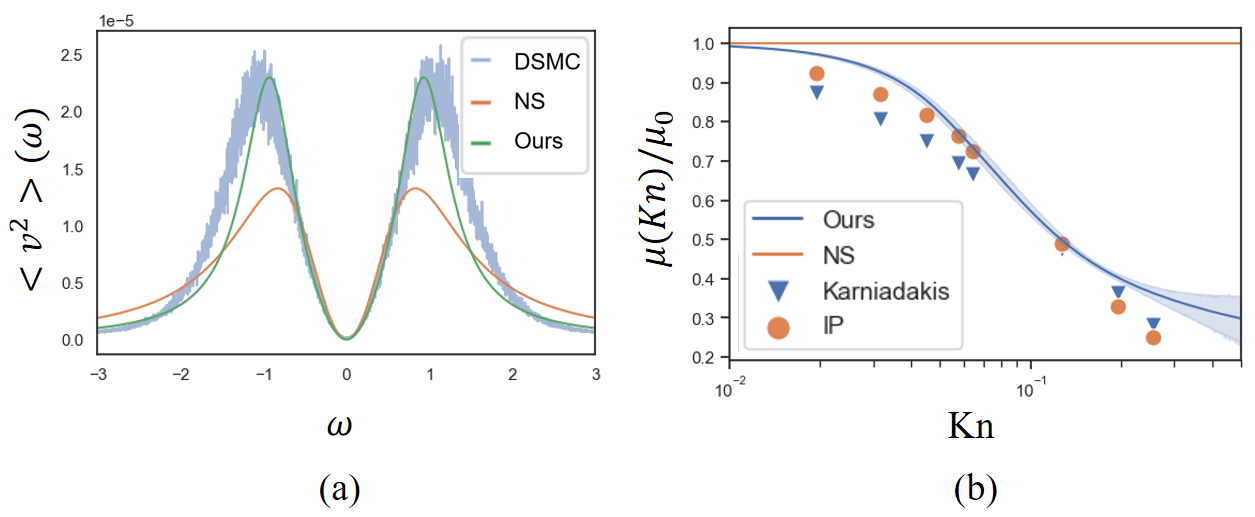}% Here is how to import EPS art
\caption{(a) \label{velocity} The test comparison between our model and NS equation predicted velocity fluctuation spectra with DSMC results at Knudsen number $Kn = 0.15$. Though our model has never trained on the velocity fluctuation spectra, it still outperforms the NS equation. (b) The effective viscosity $\frac{\mu(\Kn)}{\mu_0}$ and its $95\%$ confidence interval (shadowed) of our model for Rayleigh scattering, compared with the NS equation and results from nonlinear microchannel flow (IP, Karniadakis). To compare microchannel flow and Rayleigh scattering results, we match their Knudsen number at effective viscosity 0.5 by multiplying a constant to microchannel flow Knudsen number.  }
\end{figure}

We test the generalization ability of our model performance by predicting velocity spectra. The generalization ability ensures our model learns the rarefied dynamic physics rather than being forced to reproduce the DSMC density spectra data. As a linear benchmark, we predict the velocity fluctuation spectra of rarefied gas. Our model predicted these spectra in Fig \ref{velocity}(a), which matches with DSMC result much better than the NS equation. Moreover, to demonstrate the robustness of our model, we also plotted the $95\%$ confidence interval in Fig \ref{velocity}(b), estimated using multiple runs on randomly sampled training data. We claim our model has a robust generalization ability for rarefied gas fluctuations based on these benchmarks.

The potential risk of our model overfitting the density fluctuation spectra in training data is negligible. Overfitting refers to the phenomenon that the neural network is too powerful to remember the exact shape of spectra. However, given a Knudsen number, our neural network only models the viscosity scaling law, whose output is a number. Such a number is not enough to record the exact shape of spectra, which is a function of $\omega$. Hence it is impossible for our neural network to learn the spectra by rote, making its potential risk of overfitting negligible. However, the negligible risk of overfitting does not mean our model generalizes well to all situations.

As we will discuss later, our model does not generalize to boundary regions. We demonstrate this by comparing our model with results on microchannel flow \cite{Roohi2009ExtendingTN, Karniadakis2001MicroflowsAN} in Fig \ref{velocity}(b). The effective viscosity $\frac{\mu(\Kn)}{\mu_0}$ of our model has a similar trend compared with microchannel flow results. However, it deviates from microchannel flow even at small Knudsen numbers. The reason is microchannel flow contain boundary region with large flow property gradients. Flow properties in such boundary regions are not governed by \eqref{Non-dim Macroscopic Equations} and do not admit a Fourier decomposition. Therefore our model could not describe flows in boundary regions.

\section{Discussion} 
Data-driven models modeling physics systems typically learn PDEs consisting of derivatives of various orders \cite{Raissi2017, Rudy2017, Schaeffer2017, Raissi2019, Long2018, Long2019}. The general form of data-driven models for linear constitutive relation is a linear regression of all derivatives as in \eqref{The Linear Model on Derivatives: stress_and_flux general}. We have given it a clear physical explanation by pointing out the equivalence between constitutive relation and scaling law for transport coefficients. Our discussion also reveals that high-order derivatives enable constitutive relations to model more accurate scaling laws. The reason is additional terms of high-order derivatives in \eqref{The Linear Model on Derivatives: stress_and_flux general} contribute additional polynomials terms to \eqref{AP_The Linear Model on Derivatives: stress_and_flux, Fourier} (Appendix \ref{Ap: B}). These additional terms make scaling laws in \eqref{The Linear Model on Derivatives: stress_and_flux, not flux, Fourier} more flexible and hence more accurate. Therefore scaling laws helped us explain how high-order derivatives contribute to constitutive relations.

Scaling laws not only gives a physical explanation but also helps to avoid practical problems in learning the constitutive relation. Instead of regression on derivatives, we suggest directly modeling the scaling functions. It helps to avoid two major problems in regression on derivatives: derivative estimation and variable selection. Derivative estimation encounter stability and accuracy issue for high-order derivatives. Directly modeling the scaling functions avoids this without undermining its flexibility. Variable selection from infinite coefficients in \eqref{The Linear Model on Derivatives: stress_and_flux general} is difficult even if sparsity methods are involved. Modeling scaling functions replaced these coefficients with neural networks. In total, our argument suggests a better formulation for data-driven modeling.

Our model is implicitly connected with the Grad 13 method while avoiding its shortcomings by learning from data. Our model implicitly relies on the Grad 13 one-particle distribution $f(\mathbf{c})$ for gas molecule
\begin{equation}\label{f}
f = f_0\left(1+\frac{3}{4}\mbox{Kn} ( {c}_i  {c}_j -\frac{1}{3}\delta_{ij}\mathbf{ {c}}^2) {\sigma}_{ij}  +\frac{\mbox{Kn}}{\mbox{Pr}} (\frac{ {\mathbf{c}}^2}{2}-\frac{5}{2}) {c}_i   {q}_{i} \right)
\end{equation}
where $\mathbf{{c}}$ is the non-dimensional peculiar velocity, and $f_0$ is the Maxwell distribution. It is because this distribution is the most probable form that admits arbitrary stress ${\sigma}$ and heat flux ${q}$ as its moments, which is the pre-requisite of modeling stress and heat flux as in \eqref{The Linear Model on Derivatives: stress_and_flux general}. However, as we have shown in the result section, our model outperforms the NS equation and the Grad 13 method using only three conservation laws. The reason is our model uses the data-driven approach that learns the rarefied gas dynamics from data points equally, which converges uniformly for various Knudsen numbers. Contrarily, higher-order perturbation or moment expansion benefits little for more accurate spectra at large Knudsen numbers \cite{Wu2020}, limited by their slow convergence rate at large Knudsen numbers. In conclusion, by learning uniformly from data, the data-driven approach has demonstrated a clear advantage over the traditional Chapman-Enskog expansion and Grad's moment method in handling rarefied gas dynamics.

The implicit connection with the Grad 13 method also reveals the limitation of our model. Distribution \eqref{f} is unsuitable for strong non-Maxwellian dynamics, such as shock waves \cite{Grad1952ThePO}. Counter-intuitively, a data-driven model requires constraints rather than flexibility to resolve this, because strong non-Maxwellian dynamics tend to have correlated stress and heat flux that \eqref{f} can not handle, due to irregularly shaped distribution. Proper constraints on such correlation may be a future direction for data-driven strong non-Maxwellian models.

Similarly, our model relates to the Hilbert-Chapman-Enskog expansion hence bearing the same weakness at boundary regions. Theoretically, we could determine the coefficients in the constitutive relation model \eqref{The Linear Model on Derivatives: stress_and_flux general} by the Hilbert-Chapman-Enskog expansion. Therefore our model could be treated as a reformulation of it with data-driven enhanced convergence. However, the Hilbert-Chapman-Enskog expansion fails in boundary regions where the solutions have large gradients, such as boundary, shock, and initial layers, because the residual of expansion is proportional to gradients \cite{Cercignani1988TheBE52}. Hence our model also fails in such regions. Extending our model to boundary regions demands solving additional connection problems \cite{Cercignani1988TheBE55} concerning the gradients of gas flow, which changes the system equation. We expect those gradients to affect the scaling laws of transport coefficients by breaking the homogenous symmetry. Correspondingly, the neural network will no longer be an even function on Knudsen numbers. Flow gradients may further introduce non-local effects into scaling laws. Therefore extra equations describing the such non-local effects of transport coefficients may be required to extend our model to boundary regions.

Finally, we clarify our model's valid scenario from machine learning's point of view. Similar to other data-driven approaches, the training data scope limits our model's viability. Specifically, the limitation is in two aspects: the range of Knudsen numbers and the governing equation. Our model could only capture physics within the Knudsen number range of the training data. In our case, it covers Knudsen numbers from 0 to 0.25. Our model is unreliable outside this Knudsen number range since it extrapolates the data. This limitation on the Knudsen number range does not undermine the utility of our model because we only need to train the model once on the desired Knudsen number range before applying it to other physical scenarios. As for the governing equation, our model works only for physical scenarios governed by the system equation \eqref{Non-dim Macroscopic Equations}. However, it admits different physical scenarios correspond to distinct initial conditions, such as velocity fluctuation in our test data.

In summary of the paper, we argued that the data-driven regression models for constitutive relations are equivalent to length scaling laws of transport coefficients. Our argument not only provides a theoretical justification for data-driven models but also helps to avoid practical problems. We further modeled constitutive relations based on our argumentation. On calculating the Rayleigh scattering spectra, our model significantly outperforms the Chapman-Enskog expansion and Grad moment methods. Our argumentation also reveals the implicit assumption and limitation of data-driven constitutive relations. Further constraints and modifications are necessary for it to accommodate strong non-equilibrium dynamics and boundary layers.

\begin{acknowledgments}
We wish to acknowledge Professor Lei Wu (Southern University of Science and Technology) in offering suggestions on Rayleigh scattering modeling. We thank Ms Yuan Lan (Hong Kong University of Science and Techonology) for comments on the manuscript.
\end{acknowledgments}

\appendix
\section{The Governing Equation and Its Non-Dimensionalization} \label{Ap A}

In this section, we hydrodynamics governing equations. Then we list the detailed non-dimensionalization for these conservation laws.

The hydrodynamics equations governing the macroscopic dynamics of gases are equations of gas' statistical quantities, such as the number density $n$, mass density $\rho = m n$, the velocities $\mathbf{v}$, temperature $T$, stresses $\sigma$, heat fluxes $q$, etc. The most fundamental hydrodynamics equations are the conservation laws for mass, momentum, and energy, as described in \cite{Kardar2007StatisticalPOC37}. In our paper, we consider linearized hydrodynamics equations for one-dimensional gas. These equations describe small fluctuations of statistical quantities around a specific equilibrium state of stationary gas with density $\rho_0$ and temperature $T_0$. Specifically, one could obtain the linearized conservation laws via the first-order expansion of gases' density, velocity, and temperature at the equilibrium. Here we omit the details of the expansion and give the linearized system directly as

\begin{equation}\label{AP_Non-dim Macroscopic Equations}
	\begin{split}
		\frac{\partial  \bar{\rho}}{\partial  \bar{t}} + \frac{\partial  \bar{v}}{\partial  \bar{x}}     &= 0\\
		\frac{\partial  \bar{v}}{\partial  \bar{t}} + \frac{\partial  \bar{T}}{\partial  \bar{x}} + \frac{\partial  \bar{\rho}}{\partial  \bar{x}} &=   -\mbox{Kn}  \frac{\partial   \bar{\sigma}}{\partial  \bar{x}}  \\
		\frac{3 }{2}\frac{\partial  \bar{T}}{\partial  \bar{t}} + \frac{\partial  \bar{\rho}}{\partial  \bar{x}} &= -  \frac{15 }{4 } \mbox{Kn}\frac{\partial   \bar{q}}{\partial  \bar{x}},
	\end{split}
\end{equation} 
in which we use quantities with bars to represent the non-dimensionalized quantities. 

In the main text, we omitted bars for the simplicity of notations. We also use non-dimensionalized quantities with bar omitted in appendix \ref{Ap: B} and appendix \ref{Ap: D}. However, we use dimensionalized quantities in SI unit in appendix \ref{Ap: C}, \ref{Ap: E}, \ref{AP: F} to simplify the computation and ensure the consistency with reference.

Now we describe the details of the non-dimensionalization. Suppose we aims to compute properties of the Fourier mode of fluctuations with wavenumber $k_0$. In the non-dimensionalization process, it is natual to set the reference length scale as the wavelength $L = \frac{2\pi}{|k_0|}$ of the Fourier mode. Correspondingly, the reference time scale $T = \frac{L}{\sqrt{k_BT_0/m}} $ is time used by the sound wave to travel the distance of the reference length scale. We denote explicitly the non-dimensionalization of other quantities here. The non-dimensionalized time and spatial position in $x$ direction are $\bar{t} = \frac{t}{T}$, and $\bar{x} = \frac{x}{L}$. The non-dimensionalized Fourier wavenumber and angular frequency are $\bar{k} = k L = \frac{2\pi k}{|k_0|}$, $\bar{\omega} = \omega T$. The non-dimensionalized velocity ($x$ direction component), density, and temperature are $\bar{v}_x = \frac{T v_x}{L}$, $\bar{\rho} = \frac{\rho - \rho_0}{\rho_0}$, $\bar{T} = \frac{T-T_0}{T_0}$. Moreover, the non-dimensionalized stress and heat flux in $x$ direction are $\bar{\sigma}_{xx} =\frac{L}{\mu_0} \sqrt{\frac{m }{ k_B T_0}} \sigma_{xx}$, $\bar{q}_{x} = \frac{L}{\kappa_0 T_0 }  q_x$, in which $\mu_0$ and $\kappa_0$ are the viscosity and heat conduction coefficients at equilibrium satisfying the relations $\kappa_0 = \frac{5 k_B}{2m} \frac{\mu_0}{\mbox{Pr}}$, $\mbox{Pr}=\frac{2}{3}$. 

We define the Knudsen number $\Kn$ in \eqref{AP_Non-dim Macroscopic Equations} of the Fourier mode of fluctuations with wavenumber $k_0$ as
\begin{equation} \label{AP: non-dim kn}
\begin{split}
\Kn &=\frac{l}{2\pi/|k_0|}  \\
%\Kn(\bar{k}) &=\frac{l}{2\pi/|k|}=\frac{  |\bar{k}|}{2\pi }\Kn\\
l &= \sqrt{\frac{m}{k_B T_0}}\frac{\mu_0}{\rho_0},\\
\end{split}
\end{equation}
in which $l$ is the mean free path and $\mu_0, \kappa_0$ are the viscosity and heat conduction coefficient of the gas at equilibrium. For other Fourier modes non-dimensionalized wavenumber $\bar{k}$, their Knudsen number $\Kn(\bar{k})$ are proportional to $\bar{k}$ as follows

\begin{equation} \label{AP: non-dim kn prop}
\begin{split} \Kn(\bar{k})&=\frac{l}{2\pi/|k|}=\frac{  |\bar{k}|}{2\pi }\Kn
\end{split}
\end{equation}

Finally, we give the formula to change the reference length scale. It is helpful to change the wavenumber $k_0$ of interest to another wavenumber, which corresponds to changeing the reference length scale $L$ to the wavelength of another Fourier mode. Suppose a non-dimensionalized physical quantity $\bar{f}$ are obtained from its dimensionalized version $f$ by $\bar{f}(\bar{x}) = \frac{1}{D}f(\bar{x} L)$. Define the spatial Fourier transformation of $\bar{f}(\bar{x})$ as $\bar{g}(\bar{k})$, while the spatial Fourier transformation of $f(x)$ as $g(k)$. Here we use the Fourier transformation in the symmetrical form \cite[Eq 13.5]{riley_hobson_bence_2006}. Then  $\bar{g}(\bar{k})$ and $g(k)$ satisfies

\begin{equation} \label{AP: Fourier change scale}
\bar{g}(\bar{k}) = \frac{1}{DL} g(\frac{\bar{k}}{L})
\end{equation}
With the help of \eqref{AP: Fourier change scale}, we could easily handle Fourier transformation from one non-dimensionalization of reference length scale to other reference length scales.

\section{The Equivalence Between Data-Driven Constitutive Relations and Scaling Laws}\label{Ap: B}

This section shows the equivalence between \eqref{The Linear Model on Derivatives: stress_and_flux general} and \eqref{The Linear Model on Derivatives: stress_and_flux, not flux, Fourier} in the main text. Before the derivation, we first discuss if the constitutive relation \eqref{The Linear Model on Derivatives: stress_and_flux general} is well-defined via dimension analysis.

As shown in the non-dimensionalization process in Appendix \ref{Ap A}, our system's only degree of freedom is the Knudsen number. As a result, any non-dimensional numbers, such as the coefficients $a, b, c, d, e, f$ in the data-driven constitutive relation model \eqref{The Linear Model on Derivatives: stress_and_flux general} are functions depending on the Knudsen number which complex our analysis. 
For simplicity, in this section, we use the non-dimensionalization with $k_0 = \frac{2\pi}{l}$ which makes $L = l$. Under this non-dimensionalization, the Knudsen number of the system is fixed to $\Kn=1$, while the non-dimensionalized wavenumber $k$ corresponds to the `relative' Knudsen number of each Fourier mode.  This non-dimensionalization makes the constitutive relation

\begin{equation} \label{AP_The Linear Model on Derivatives: stress_and_flux general}
	\begin{split}
		 {\sigma} &= - \sum_{n=1}^{\infty}  \left( {a}_n \frac{\partial^n  {v}}{\partial  {x}^n}  + {c}_n \frac{\partial^n  {\rho}}{\partial  {x}^n} +{e}_n \frac{\partial^n  {T}}{\partial  {x}^n}\right) \\
		 {q} &=-\sum_{n=1}^{\infty} \left(  {b}_n\frac{\partial^n  {T}}{\partial  {x}^n}+{d}_n \frac{\partial^n  {\rho}}{\partial  {x}^n}+{f}_n \frac{\partial^n  {v}}{\partial  {x}^n} \right)
	\end{split}
\end{equation}
well-defined with coefficients $a, b, c, d, e, f$ independent of the global $\Kn$. Note that this constitutive relation is exactly \eqref{The Linear Model on Derivatives: stress_and_flux general} in the main text.

We now show that the above data-driven constitutive relation \eqref{AP_The Linear Model on Derivatives: stress_and_flux general}
is equivalent to the constitutive relation with scaling laws \eqref{The Linear Model on Derivatives: stress_and_flux, not flux, Fourier} in the form
\begin{equation} \label{AP_The Linear Model on Derivatives: stress_and_flux, not flux, Fourier}
	\begin{split}
		  \tilde{\sigma}( k) & = -i k \frac{4}{3} \frac{\mu(k)}{ \mu_0} \tilde{v}( k); \quad \mu( k) \ge 0 \\
		 \tilde{q}(  k) &=-i k \frac{\kappa(k)}{\kappa_0} \tilde{T}( k); \quad \kappa(k) \ge 0 ,
	\end{split}
\end{equation}
The constraint on entropy production is the key to achieving the equivalence. According to \cite{Landau1980hf}, The total entropy production rate of the conservation laws of the mass, moment, and energy is
\begin{equation} \label{AP_Hydrodynamic Fluctuations:Total entropy change rate, expand 1}
\begin{split}
	\dot{S} & = \int -\frac{q_j}{T^2}\partial_j T - \frac{\sigma_{ij}}{2T} \left(  \frac{\partial v_i}{\partial x_j} + \frac{\partial v_j}{\partial x_i}  \right) d^3 x,
\end{split}
\end{equation} 
, in which we use the Einstein's summation convention with $i,j \in \{x,y,z\}$ as the dummy indices. As a fundamental constraint, the second law of thermodynamics requires the total entropy change rate $\dot{S} \ge 0$. However, it is not enough to deduce the equivalence we desired.

Linearization of the system dramatically helps us by simplifying the constraints. First, we could replace the temperature $T$ with the equilibrium temperature $T_0$ as a first-order approximation since $\delta T = T-T_0$ is a small quantity. 
\begin{equation} \label{AP_Hydrodynamic Fluctuations:Total entropy change rate, expand 2}
\begin{split}
	\dot{S} & \approx \int -\frac{q_j}{T_0^2}\partial_j T - \frac{\sigma_{ij}}{2T_0} \left(  \frac{\partial v_i}{\partial x_j} + \frac{\partial v_j}{\partial x_i}  \right) d^3 x \ge 0,
\end{split}
\end{equation} 
Second, the two terms for velocity and temperature field must be non-negative separately
\begin{equation} \label{AP_The DSMC Result:  with linear model: Entropy condition}
 \int q_j\partial_j T d\bold{x} \le 0; \quad  \int\sigma_{ij} \left(  \frac{\partial v_i}{\partial x_j} +  \frac{\partial v_j}{\partial x_i}  \right)  d\bold{x} \le 0,
\end{equation}
because we know from statistical mechanics that the deviations of velocity and temperature from equilibrium are statistically independent \cite{Landau1980ipt}. If we reduce our problem to the 1D case and denote $v_x$ as $v$, $\sigma_{xx}$ as $\sigma$, and $q_x$ as $q$. A Fourier transformation on \eqref{AP_The DSMC Result:  with linear model: Entropy condition} with the help of the Paseval's theorem lead us to
\begin{equation} \label{AP_The DSMC Result:  with linear model: Entropy condition, 1D, Fourier int}
	 \int \tilde{q}(k) (i k\tilde{T}(k))^* dk\le 0; \quad  \int \tilde{\sigma}(k) (i k \tilde{v}(k))^* dk \le 0,
\end{equation}
in which symbol with tilde represents the Fourier transform of corresponding function in $x$ and $*$ indicates complex conjugate. Finally, the fluctuation of statistical quantities generally behaves like white noise that spreads over the entire spectra. Therefore the linear system should be stable for each wavenumber $k$, which means the entropy production from each wavenumber must be non-negative.
\begin{equation} \label{AP_The DSMC Result:  with linear model: Entropy condition, 1D, Fourier}
	  \tilde{q}(k) (i k\tilde{T}(k))^*\le 0; \quad   \tilde{\sigma}(k) (i k \tilde{v}(k))^* \le 0,
\end{equation}

%Furthermore, we assume that the model system is in quasi-equilibrium to remove the integration in \eqref{AP_The DSMC Result:  with linear model: Entropy condition, 1D, Fourier int}. Specifically, quasi-equilibrium means the second law of thermodynamics applies to each volume element $d^3 x$
%\begin{equation} \label{AP_Hydrodynamic Fluctuations:entropy change rate constraint}
%\begin{split}
%	 \left(-\frac{q_j}{T}\partial_j T - \frac{\sigma_{ij}}{2} \left(  \frac{\partial v_i}{\partial x_j} + \frac{\partial v_j}{\partial x_i}  \right) \right)d^3 x \ge 0.
%\end{split}
%\end{equation} 
%This assumption implies that we model rarefied gas dynamics using some 'model particles' with negligible mean-free paths $s$. Particularly, it means that the constitutive relation \eqref{AP_The Linear Model on Derivatives: stress_and_flux general} describes the dynamics of the 'model particles' that behave similarly to rarefied gas.
%In this case, the entropy exchange between volume elements only happens at their surface, occupying the volume $s (dx)^2$. Such volume is negligible compared to the volume $(dx)^3$ of the volume element. Therefore each volume element could be treated as a quasi-closed system with the constraint \eqref{AP_Hydrodynamic Fluctuations:entropy change rate constraint} valid. 

Now we discuss what constraint \eqref{AP_The DSMC Result:  with linear model: Entropy condition, 1D, Fourier} imposes on the constitutive relations. First, stress $\sigma$ depends on the velocity field $v$ only, while the heat flux $q$ depends solely on the temperature field $T$. This is the only way to ensure \eqref{AP_The DSMC Result:  with linear model: Entropy condition, 1D, Fourier} since density $\rho$, velocity $v$, and temperature $T$ fluctuations are statistically independent \cite{Landau1980indp, Landau1980ipt} with no guarantee in their mutual products. Therefore the linear constitutive relations \eqref{AP_The Linear Model on Derivatives: stress_and_flux general} reduce to the from \begin{equation} \label{AP_The Linear Model on Derivatives: stress_and_flux}
	\begin{split}
		 {\sigma} &= - \sum_{n=1}^{\infty}   {a}_n \frac{\partial^n  {v}_x}{\partial  {x}^n}  \\
		 {q} &=-\sum_{n=1}^{\infty}  {b}_n\frac{\partial^n  {T}}{\partial  {x}^n} ,
	\end{split}
\end{equation} under the condition of non-decreasing entropy. 
%Its non-dimensionalized form according to the definitions in \ref{Ap A} is
%\begin{equation} \label{AP_The Linear Model on Derivatives: stress_and_flux no dim}
%	\begin{split}
%		\bar{\sigma} &= - \sum_{n=1}^{\infty}  \frac{a_n}{\mu \Delta x^{n-1}} \frac{\partial^n \bar{v}_x}{\partial \bar{x}^n} = - \sum_{n=1}^{\infty}  \bar{a}_n \frac{\partial^n \bar{v}_x}{\partial \bar{x}^n}  \\
%		\bar{q} &=-\sum_{n=1}^{q} \frac{b_n}{\kappa \Delta x^{n-1}} \frac{\partial^n \bar{T}}{\partial \bar{x}^n}=-\sum_{n=1}^{q} \bar{b}_n\frac{\partial^n \bar{T}}{\partial \bar{x}^n} ,
%	\end{split}
%\end{equation}
%in which $\bar{a}_n =  \frac{a_n}{\mu \Delta x^{n-1}}$, $\bar{b}_n = \frac{b_n}{\kappa \Delta x^{n-1}}$. 
A Fourier transformation of the above forms gives 
\begin{equation} \label{AP_The Linear Model on Derivatives: stress_and_flux no dim Fourier}
	\begin{split}
	\tilde{\sigma}(k) & = - \sum_{n=1}^{\infty}  a_n (i k)^n v(k)  \\
		\tilde{q}(k) &=-\sum_{n=1}^{q} b_n(i k)^n T(k) ,
	\end{split}
\end{equation}

Combine it with the constraint \ref{AP_The DSMC Result:  with linear model: Entropy condition, 1D, Fourier}, leads us to
\begin{equation} \label{AP_The Linear Model on Derivatives: stress_and_flux entropy}
	\begin{split}
		\sum_{n=1}^{\infty} (i k)^{n+1} a_n \left|\tilde{v}(k)\right|^2 &\le 0 \\
		\sum_{n=1}^{\infty} (i k)^{n+1} b_n \left|\tilde{T}(k)\right|^2 &\le 0  \\
	\end{split}
\end{equation}
There should not be any imaginary part appearing in the LHS of \eqref{AP_The Linear Model on Derivatives: stress_and_flux entropy}. Hence all terms with odd powers on $ik$ vanishes. What remains is the following
\begin{equation} \label{AP_The Linear Model on Derivatives: stress_and_flux entropy, explicit terms}
	\begin{split}
 (k^2 a_1 - k^4 a_3 + k^6 a_5 - k^8 a_7 \cdots ) \left|\tilde{v}(k)\right|^2 &\ge 0 \\
 (k^2 b_1 - k^4 b_3 + k^6 b_5 - k^8 b_7 \cdots )\left|\tilde{T}(k)\right|^2&\ge 0  \\
 a_0 = a_2 = \cdots = a_{{2n}} = \cdots= &0 \\
  b_0 = b_2 = \cdots = b_{{2n}} = \cdots= &0 \\
	\end{split}
\end{equation}
Substitute \eqref{AP_The Linear Model on Derivatives: stress_and_flux entropy, explicit terms} into \eqref{AP_The Linear Model on Derivatives: stress_and_flux no dim Fourier} and take derivative w.r.t $x$ gives us the constitutive relation in Fourier space
\begin{equation} \label{AP_The Linear Model on Derivatives: stress_and_flux, Fourier}
	\begin{split}
		\widetilde{\partial_{ {x}}  {\sigma}}( {k}) &= ( {k}^2  {a}_1 -  {k}^4  {a}_3 +  {k}^6  {a}_5  \cdots )  \tilde{v}_{ {k}} =  {k}^2 M( {k})\tilde{v}{ (k)}\\
		\widetilde{\partial_{ {x}}  {q}}( {k}) &= ( {k}^2  {b}_1 -  {k}^4  {b}_3 +  {k}^6  {b}_5  \cdots ) \tilde{T}_{ {k}}=  {k}^2 K( {k})\tilde{T}{ (k)} , \\
	\end{split}
\end{equation}
in which function $M,K$ are the infinite sum of series and should be non-negative even functions of $k$. Note that the Knudsen number $\Kn(k)$ corresponds to the Fourier mode with wavenumber $k$ is $Kn(k) = \frac{|k|}{2\pi}$ according to \eqref{AP: non-dim kn}. Therefore functions $M,K$ are even functions of Knudsen numbers, hence are well defined in the sense of dimensional analysis. 

Functions $M,K$ are closely related to viscosity and heat conduction coefficients. They could be rewritten in the following form
\begin{equation} \label{AP_The Linear Model on Derivatives: stress_and_flux, not flux, Fourier MN,appendix}
	\begin{split}
		 M({k}) &= \frac{4}{3}\frac{\mu({k})}{\mu_0}\\
		 K({k}) &= \frac{\kappa({k})}{\kappa_0}\\
	\end{split}
\end{equation}
in which $\mu({k}),\kappa({k})$ are scaling laws of viscosity and heat conduction coefficients satisfying $\mu(0) = \mu_0$ and $\kappa(0) = \kappa_0$. Under this notation, $\mu({k}) =\mu_0 ,\kappa({k}) =\kappa_0$ exactly corresponds to the constitutive relation for the NS equation.

We could further deduce the stress and heat flux under the same notation with \eqref{AP_The Linear Model on Derivatives: stress_and_flux, not flux, Fourier MN,appendix} by removing the spatial derivative in \eqref{AP_The Linear Model on Derivatives: stress_and_flux, Fourier}. The result is exactly \eqref{AP_The Linear Model on Derivatives: stress_and_flux, not flux, Fourier}, which completes our derivation from \eqref{AP_The Linear Model on Derivatives: stress_and_flux general} to \eqref{AP_The Linear Model on Derivatives: stress_and_flux, not flux, Fourier}. Hence we have shown the equivalence between \eqref{The Linear Model on Derivatives: stress_and_flux general} and \eqref{The Linear Model on Derivatives: stress_and_flux, not flux, Fourier} in the main text.

%We could further deduce the stress and heat flux by removing the spatial derivative:
%\begin{equation} \label{AP_The Linear Model on Derivatives: stress_and_flux, not flux, Fourier,appendix}
%	\begin{split}
%		 {\sigma}({k}) & = -i {k} M({k})\tilde{v}_{{k}} \quad M({k}) \ge 0 \\
%		{q}({k}) &=-i {k} K({k})\tilde{T}_{{k}} \quad K({k}) \ge 0  . \\
%	\end{split}
%\end{equation}

In addition, we introduce the constitutive relation under the same non-dimensionalization with Appendix $\ref{Ap A}$, which is useful in the computation of spectra. Recall that we have used the reference length scale $L=l$ in this section instead of $L=\frac{2\pi}{|k_0|}$ in Appendix $\ref{Ap A}$. If we change the reference length scale to $L=\frac{2\pi}{|k_0|}$. The constitutive relation \eqref{AP_The Linear Model on Derivatives: stress_and_flux, Fourier} will becomes 
\begin{equation} \label{AP_The Linear Model on Derivatives: stress_and_flux, Fourier rescale}
	\begin{split}
		\widetilde{\partial_{ {x}}  {\sigma}}( {k}) &=  {k}^2 M( {k}\Kn)\tilde{v}{ (k)}\\
		\widetilde{\partial_{ {x}}  {q}}( {k}) &=  {k}^2 K( {k}\Kn)\tilde{T}{ (k)} , \\
	\end{split}
\end{equation}
in which $\Kn = \frac{l}{2\pi/|k_0|}$. This result could be derived with the help of \eqref{AP: Fourier change scale}. The merit of choosing $L=\frac{2\pi}{|k_0|}$ as the reference length scale is that the non-dimensionalized $k=2\pi$ exactly corresponds to the dimensionalized wavenumber $k_0$. Therefore we could substitute $k$ with $2\pi$ everywhere if we are only interested in the dynamics at wavenumber $k_0$.

\section{Rayleigh Scattering and Density Fluctuations} \label{Ap: C}
This section we introduce the Rayleigh scattering spectra and show that it is proportional to the density fluctuation spectra. The Rayleigh scattering was discovered by Lord Rayleigh back in the nineteenth century. It is the reason for the blue color of the sky in daytime and twilight. Specifically, the Rayleigh scattering is due to the refraction of electromagnetic waves (EM waves) passing through media with density fluctuations. Such fluctuation leads to changes in the dielectric constant hence generating refracted EM waves. This section only gives a rough introduction emphasizing the physical picture of Rayleigh scattering and its connection to density fluctuation spectra. One could refer to \cite{Pecora1964, jackson1999classical, landau2013electrodynamics} for a detailed treatment of the Rayleigh scattering spectra. In addition, we do not use non-dimensionalization in this section for simplicity in discussing the related electrodynamics.

The {\em Rayleigh scattering} describes the refraction of incident electromagnetic waves  passing through gas media with stochastic density fluctuation  $\delta \rho$. Considering the incident electromagnetic wave as plain EM wave with given wavevector $\mathbf{k}_i$,
\begin{equation} \label{AP_Rayleigh Scattering Spectra: Expansion, 0th solution}
\begin{split}
\mathbf{E}_{inc} &= \boldsymbol{\xi}_0 \exp (i \mathbf{k}_i \cdot \mathbf{r} + i \omega_i t) \\
\left| \mathbf{k}_i\right|&=\frac{\sqrt{\epsilon_{0}} \omega_i}{c}
\end{split}
\end{equation}
With $\boldsymbol{\xi}_0 $ be the polarization vector, $\mathbf{k}_i$ the incident wave vector, $\omega_i$ the incident wave frequency, $c$ is the speed of light in vacuum, and $\epsilon_0$ is the dielectric constant of gas. The propagation of the incident wave is governed by the Maxwell's equations in matter without source \cite[Eq 10.21]{Jackson1999ClassicalE3}. We approximate the permeability $\mu$ of gas with the permeability of the vacuum $\mu_0$ since they are very close for most materials. Under this approximation, the Maxwell's equations reduces to
\begin{equation} \label{AP: Rayleigh Scattering Spectra: Maxwell Equation, rearange}
	\begin{split}
		\nabla \cdot \bold{D} &= 0 \\
		\nabla \times\nabla \times \bold{E} &= -\frac{1}{c^2}\frac{\partial^2 \bold{D}}{\partial t^2},\\
	\end{split}
\end{equation}
in which $\bold{D} = \epsilon\bold{E}$, $\epsilon$ is the dielectric constant of gases. This equation governs the propagation of the incident wave in gas media.

Now we analyze how the stochastic density fluctuation  $\delta \rho$ affects the propagation of the incident wave $\mathbf{E}_{inc}$. The dielectric constant of gases $\epsilon$ is a known function of the gas density. Therefore small fluctuations $\delta \rho$ in the gas density leads to the perturbation in $\epsilon$ and $\bold{E}$
\begin{equation} \label{AP_Rayleigh Scattering Spectra: Expansion}
	\begin{split}
\epsilon(\bold{r},t) &=\epsilon_{0} + \epsilon_{1}(\bold{r},t) + \cdots;\  \epsilon_{1}(\bold{r},t) =\frac{\partial \epsilon}{\partial \rho}(\rho_0) \delta \rho(\bold{r},t)   \\
\bold{E}(\bold{r},t) &= \bold{E}_{inc}(\bold{r},t) + \bold{E}_1(\bold{r},t) + \cdots, \\
	\end{split}
\end{equation}
in which $\rho_0$ is the equilibrium density of gas. Substitute the above expansion in to \eqref{AP: Rayleigh Scattering Spectra: Maxwell Equation, rearange} yields perturbation equation of different orders obtained via perturbation theory in \cite{Pecora1964}. Specifically, the first order perturbation of the electric field $\bold{E}_1$ could be calculated by solving the following Helmholtz equation
\begin{equation} \label{Rayleigh Scattering Spectra: Expansion, 1th, rearange}
		\nabla^2 \bold{D}_1-\frac{\epsilon_{0}}{c^2}\frac{\partial^2 \bold{D}_1}{\partial t^2} = -\nabla \times\nabla \times (\epsilon_{1} \bold{E}_{inc})
\end{equation}
in which $\bold{D}_1 = \epsilon_{0}\bold{E}_1 +\epsilon_{1} \bold{E}_0$. 

Under specific scenarios,  we could find analytical solutions to \eqref{Rayleigh Scattering Spectra: Expansion, 1th, rearange} with the help of the Born approximation. Specifically, we consider observing the EM wave at position $\bold{r}$ scattered from gases of a certain volume $V$ centered at the origin. In addition, we assume $\epsilon_1 = \delta \rho = 0$ outside the volume $V$. The distance between the observation position $\bold{r}$ and the volume $V$ is so large compared to the radius of the volume $V$ that it allows us to adopt the Born approximation \cite[Eq 117.4]{landau2013electrodynamics} \cite[Eq 10.53, 10.73]{griffiths_schroeter_2018} to compute the electric field
\begin{equation} \label{Rayleigh Scattering Spectra: Expansion, 1th, e field}
\begin{split}
	\bold{E}_1(\bold{r}, \omega_f) &=  \frac{-\hat{\bold{r}}\times\hat{\bold{r}}\times\boldsymbol{\xi}_0}{ \sqrt{2/\pi}  c^2} \frac{\omega_f^2 e^{i k_f r}}{r} \tilde{\epsilon}_{1} (\bold{k}_f-\bold{k}_i,\omega_f-\omega_i ) %\\&\frac{1}{4\pi}\times  \int dt \int d \bold{r}_0^3 e^{i (\bold{k}_i-\bold{k}_f)\cdot \bold{r}_0}   \epsilon_{1} (\bold{r}_0,t ) e^{i (\omega_i-\omega_f) t} 
\end{split}
\end{equation}
in which $\bold{E}_1(\bold{r}, \omega_f)$ is the Fourier transform of $\bold{E}_1(\bold{r}, t)$ w.r.t time, $k_f = \sqrt{\epsilon_{0} }\omega_f/c$, $\bold{k}_f =k_f \hat{\bold{r}}$, $\hat{\bold{r}}$ is the unit vector along the direction of $\bold{r}$, $r = |\bold{r}|$, $\tilde{\epsilon}_{1} (\bold{k},\omega)$ is the Fourier transformation of the function $\epsilon_{1}(\bold{r},t)$ w.r.t spatial and temporal coordinate. We call $\bold{E}_1$ as the electric field of scattered EM wave, whose amplitude is proportional to the perturbation of the dielectric constant.

Next, we define the Rayleigh scattering spectra as the intensity spectra of the scattered electric field $\bold{E}_1$ and connect it to the density fluctuation spectra. The Rayleigh scattering spectra refers to the intensity spectra $I$ of the scattered EM wave $\bold{E}_1$
\begin{equation} \label{AP_Rayleigh Scattering Spectra: spectral density}
    I(\bold{r},\omega_f) =  \left< \bold{E}_1^2 \right>(\bold{r}, \omega_f) = \frac{\sqrt{2\pi}}{T}|	\bold{E}_1(\bold{r}, \omega_f)|^2 \\
\end{equation}
in which we assume $\bold{E}_1$ is a periodic function in time with period $T$ and $\left< \bold{E}_1^2 \right>$ is the spectra of $\bold{E}_1$ w.r.t time. 

We explain the definition of spectra in \eqref{AP_Rayleigh Scattering Spectra: spectral density} in this paragraph. The spectra of a real function $f(t)$ with period $T$ is defined as
\begin{equation} \label{AP_Rayleigh Scattering Spectra: spectral definition}
\begin{split}
       \left< f^2 \right>(\omega)&=  \left<\mathcal{F}_t\left[f(s)f(s+t)\right]\right>_s  \\&=\frac{1}{\sqrt{2\pi}T} \int_{-\frac{T}{2}}^{\frac{T}{2}} ds \int_{-\frac{T}{2}}^{\frac{T}{2}}  dt   f(s)f(s+t) e^{-i\omega t}, \\
\end{split}
\end{equation}
in which $\left<\right>_s$ represents the ensemble average w.r.t s defined as $\left<f(s)\right>_s =\frac{1}{T} \int_{-\frac{T}{2}}^{\frac{T}{2}} f(s) ds$, $\mathcal{F}_t$ represents the Fourier transformation for periodic function w.r.t $t$ defined as $\mathcal{F}_t(f)(\omega) = \frac{1}{\sqrt{2\pi}} \int_{-\frac{T}{2}}^{\frac{T}{2}}    f(t) e^{-i\omega t} dt$. The Wiener–Kinchin theorem \cite{riley_hobson_bence_2006_autoc} simplify the definition \eqref{AP_Rayleigh Scattering Spectra: spectral definition} to

\begin{equation} \label{AP_Rayleigh Scattering Spectra: spectral definition to Fourier}
        \left< f^2 \right>(\omega)= \frac{\sqrt{2\pi}}{T}|\tilde{f}(\omega)|^2, %\frac{1}{\sqrt{2\pi}} \int_{-\infty}^{\infty} ds \int_{-\infty}^{\infty} dt   f(s)f(s+t) e^{-i\omega t}, \\
\end{equation}
where $\tilde{f}(\omega) = \mathcal{F}_t(f)(\omega)$ is the Fourier transformation of $f$. Similarly, the spectra for vector valued function $\bold{v}(t)$ with period $T$ are
\begin{equation} \label{AP_Rayleigh Scattering Spectra: spectral definition to Fourier}
\begin{split}
        \left< \bold{v}^2 \right>(\omega)&= \left<\mathcal{F}_t\left[\bold{v}(s)\cdot \bold{v}(s+t)\right]\right>_s = \frac{\sqrt{2\pi}}{T}|\tilde{\bold{v}}(\omega)|^2, %\frac{1}{\sqrt{2\pi}} \int_{-\infty}^{\infty} ds \int_{-\infty}^{\infty} dt   f(s)f(s+t) e^{-i\omega t}, \\
\end{split}
\end{equation}
which is exactly the definition we used in \eqref{AP_Rayleigh Scattering Spectra: spectral density}.

We are ready to show that the Rayleigh scattering intensity spectra \eqref{AP_Rayleigh Scattering Spectra: spectral density} is proportional to the density fluctuation spectra. The term $|	\bold{E}_1(\bold{r}, \omega_f)|^2$ in \eqref{AP_Rayleigh Scattering Spectra: spectral density} could be calculated from \eqref{Rayleigh Scattering Spectra: Expansion, 1th, e field} as

\begin{equation} \label{Rayleigh Scattering Spectra: Expansion, 1th, e field sq}
\begin{split}
	|\bold{E}_1(\bold{r}, \omega_f)|^2 &=  \frac{|\boldsymbol{\xi}_0|^2\sin(\psi)^2 \omega_f^4}{2 c^4 r^2/\pi}  |\tilde{\epsilon}_{1} (\bold{k}_f-\bold{k}_i,\omega_f-\omega_i )|^2,
\end{split}
\end{equation}
where $\psi$ is the angle between $\bold{r}$ and $\boldsymbol{\xi}_0$. The perturbation of the dielectric constant $\epsilon_{1}$ is proportional to the perturbation of density $\delta \rho$ by its definition in \eqref{AP_Rayleigh Scattering Spectra: Expansion}, hence we have
\begin{equation} \label{AP_Rayleigh Scattering Spectra: spectral density fin}
\begin{split}
       |\tilde{\epsilon}_{1} (\bold{k},\omega )|^2 &= \left(\frac{\partial \epsilon}{\partial \rho}\right)^2  |\delta\tilde{ \rho} (\bold{k}, \omega)|^2
\end{split}
\end{equation}
in which the $\delta\tilde{ \rho} (\bold{k}, \omega)$ is the Fourier transformation of $\delta \rho(\bold{r},t)$ in both the spatial and temporal coordinates. 

The equation \eqref{AP_Rayleigh Scattering Spectra: spectral density fin} establish the connection between intensity spectra \eqref{AP_Rayleigh Scattering Spectra: spectral density} and the density fluctuation spectra. Assuming $\delta \rho(\bold{r}, t)$ to be periodic in time with period T and noting that it vanishes outside volume $V$, the explicit formula for $\delta\tilde{ \rho}$ as the Fourier transformation of $\delta \rho$ is

\begin{equation} \label{AP_Spectral Resolution of fluctuations: The fourier transformation relation}
   \delta\tilde{ \rho} (\bold{k}, \omega) = \frac{1}{(2\pi)^2} \int_{-T/2}^{T/2}dt  \int_V d\bold{r}^3 \delta \rho(\bold{r}, t)  e^{-i\omega t} e^{- i\bold{k}\cdot \bold{r}},    
\end{equation}
 Note that $\delta\tilde{ \rho}$ is equal to the Fourier transform of the periodic extension of $\delta \rho$ with unit cell $V$. Therefore we could treat $\delta \rho$ as if it is a periodic function in space, hence define the density fluctuation spectra by extending \eqref{AP_Rayleigh Scattering Spectra: spectral definition to Fourier} to both the spatial and temporal coordinates

\begin{equation} \label{AP_Spectral Resolution of fluctuations: The spectra, meaning}
\begin{split}
    \left<\delta \rho^2\right>(\omega, \bold{k}) &= \frac{(2\pi)^{\frac{d+1}{2}}}{T V}|\delta\tilde{ \rho} (\bold{k}, \omega)|^2 \\
\end{split}
\end{equation}
in which $d$ is the spatial dimension of volume $V$. Finally, we get the intensity spectra $I$ of the scattered EM wave $\bold{E}_1$ in terms of the density fluctuation spectra by combining \eqref{AP_Rayleigh Scattering Spectra: spectral density} \eqref{Rayleigh Scattering Spectra: Expansion, 1th, e field sq} \eqref{AP_Rayleigh Scattering Spectra: spectral density fin} \eqref{AP_Spectral Resolution of fluctuations: The spectra, meaning}

\begin{equation} \label{Rayleigh Scattering Spectra: Expansion, 1th, e field sq final}
I(\bold{r},\omega_f) = \frac{V}{\sqrt{2\pi}^3} \frac{|\boldsymbol{\xi}_0|^2\sin(\psi)^2 \omega_f^4}{2 c^4 r^2/\pi}  \left(\frac{\partial \epsilon}{\partial \rho}\right)^2 \left<\delta \rho^2\right>(\delta \omega, \delta \bold{k}) ,
\end{equation}
in which $\delta \bold{k} = \bold{k}_f-\bold{k}_i,\delta \omega = \omega_f-\omega_i$. 

The Rayleigh scattering intensity spectra \eqref{Rayleigh Scattering Spectra: Expansion, 1th, e field sq final} follow the famous $\omega_f^4$ frequency dependence, which means blue light (high $\omega_f$) is scattered more strongly than red light (low $\omega_f$). This frequency dependence is responsible for the blue color of the sky in daytime and twilight. Moreover, the most intense scattering happens at $\psi=\frac{\pi}{2}$ when one observes the scattering light in the direction perpendicular to the incident wave. Therefore the zenith is bluer than the sky at the horizon in the daytime. However, the detailed shape of the intensity spectra $I$ proportionally depends on the density fluctuation spectra $\left<\delta \rho^2\right>$, which is determined by the hydrodynamics of gases in the scattering region.

\section{Calculating the density fluctuation spectra} \label{Ap: D}

This section computes the density fluctuation spectra $\left<\delta \rho^2\right>(\omega, k)$ introduced in the previous section. For simplicity, We omit the $\delta$ symbol throughout this section, making the notation consistent with Appendix \ref{Ap A}. Moreover, we non-dimensionalize the density fluctuation spectra as a function of the Knudsen number $\Kn$ and the non-dimensionalized $\omega$ in the form $\left< \rho^2\right>(\omega, \Kn)$. The computation largely follows \cite{Landau1980hfcal}.

First, we investigate the symmetries of spectra we shall exploit in computing the density fluctuation spectra. For a real function $f(t)$ with period $T$, its correlation function

\begin{equation} \label{AP_Rayleigh Scattering Spectra: correlation}
       \left<f^2\right>(t)=  \left<f(s)f(s+t)\right>_s
\end{equation}
is an even function satisfying $ \left<f^2\right>(-t) =  \left<f^2\right>(t)$. This could be deduced from the definition of ensemble average \eqref{AP_Rayleigh Scattering Spectra: spectral definition} in the previous section. It also holds when $T$ tends to infinity if $f$ is a stationary process. 

The spectra $\left<f^2\right>(\omega)$ defined in \eqref{AP_Rayleigh Scattering Spectra: spectral definition} is exactly the Fourier transformation of the correlation function $ \left<f^2\right>(t)$. From now on we distinguish them by denoting the spectra $\left<f^2\right>(\omega)$ as $\left<f^2\right>_\omega$ while keeping the notation $\left<f^2\right>$ for $ \left<f^2\right>(t)$. 

We could describe the spectra $\left<f^2\right>_\omega$ in terms of a one-sided Fourier transformation since the correlation function $ \left<f^2\right>(t)$ is even. Specifically, we define the one-sided Fourier transformation of $ \left<f^2\right>(t)$ as

\begin{equation} \label{AP_Spectral Resolution of fluctuations: The FT, one sided}	
   \left<f^2\right>_\omega^{(+)}(\omega) = \frac{1}{\sqrt{2\pi}}\int_0^{\infty} \left<f^2\right>(t) e^{-i \omega t} dt.
\end{equation}
It is enough to obtain the full spectra $\left<f^2\right>_\omega(\omega) $ because the even correlation function $ \left<f^2\right>(t)$ gives us the property
\begin{equation} \label{AP_Spectral Resolution of fluctuations: The FT, one sided 2}	
  \left<f^2\right>_\omega(\omega) = 2 \mbox{Re}\left[  \left<f^2\right>_\omega^{(+)}(\omega)\right] ,
\end{equation}
in which $\mbox{Re}$ represents the real part of complex numbers. Note that the  one-sided Fourier transformation we use here is just the complex version of the Fourier cosine transformation. The one-sided Fourier transformation also have the property
\begin{equation} \label{AP_Spectral Resolution of fluctuations: The FT, one sided diff}	
    \left(\partial_t g \right)_\omega^{(+)}(\omega) = i \omega g_\omega^{(+)}(\omega) - \frac{g(0)}{\sqrt{2\pi}}
\end{equation}
if $g$ vanishes at infinity.

Another important property of the correlation is that the correlation $\left<f g\right> = \left<f(s)g(s+t)\right>_s$ between periodic function $f$ and $g$ is a linear functional acting on $g$. It commute with the derivation $\partial$ w.r.t $g$

\begin{equation} \label{AP_Rayleigh Scattering Spectra: exchange }
       \left<f \partial g\right>= \partial \left<f g\right>
\end{equation}
Consequently, the stress $\sigma(v)$ and heat flux $q(T)$ as linear functionals of $v$ and $T$ in the form of \eqref{AP_The Linear Model on Derivatives: stress_and_flux} satisfies 
\begin{equation} \label{AP_Rayleigh Scattering Spectra: exchange constitutive relation}
\begin{split}
      \left<f \sigma(v)\right> &= \sigma (\left<f v\right>)\\
      \left<f q(T)\right> &= q (\left<f T\right>)
\end{split}
\end{equation}
because $\sigma,q$ are linear combinations of spatial derivatives $\partial_x$ that satisfies \eqref{AP_Rayleigh Scattering Spectra: exchange }.

With these properties, we are ready to compute the density fluctuation spectra from the linear system \eqref{AP_Non-dim Macroscopic Equations} and constitutive relation \eqref{The Linear Model on Derivatives: stress_and_flux, not flux, Fourier}. Taking the correlation between density $\rho$ and the governing equation \eqref{AP_Non-dim Macroscopic Equations}, we obtain the linear system
\begin{equation}\label{AP_Non-dim Macroscopic Equations 2}
	\begin{split}
		\frac{\partial \left< {\rho}^2 \right>}{\partial  {t}} + \frac{\partial \left< {\rho v} \right>}{\partial  {x}}     &= 0\\
		\frac{\partial  {\left< {\rho}v \right>}}{\partial  {t}} + \frac{\partial  \left<{\rho T} \right>}{\partial  {x}} + \frac{\partial  \left<{\rho^2}\right>}{\partial  {x}} &=   -\mbox{Kn}  \frac{\partial   \sigma( \left<{\rho v}\right> )}{\partial  {x}}  \\
		\frac{3 }{2}\frac{\partial  \left<{\rho T}\right>}{\partial  {t}} + \frac{\partial  \left<{\rho^2}\right>}{\partial  {x}} &= -  \frac{15 }{4 } \mbox{Kn}\frac{\partial   q(\left<{\rho T}\right>)}{\partial  {x}},
	\end{split}
\end{equation} 
The above linear system governs the correlations of density with density, velocity, and temperature. However, the initial conditions for the linear system are required to determine the correlations completely. 

The initial conditions of \eqref{AP_Non-dim Macroscopic Equations 2} describe the two-point correlations of densities, velocities, and temperatures between two simultaneous locations separated by distance $x$ at $t=0$. Such correlation vanishes if the distance $x$ is non-zero since changes in one place require time to propagate to another. Therefore initial conditions should be delta function $\delta(x)$ multiplied by some amplitude constants. These amplitude constants could be determined by the fluctuation theory in statistical mechanics. The initial condition for $ \left<  {\rho}^2 \right>(0, x)$ could be deduced from \cite[Eq 88.2]{Landau1980indp} with non-dimensionalization and DSMC's Monte Carlo effects considered. As for $ \left<  {\rho v} \right>$ and $\left<  {\rho T} \right>$, they vanishes at $t=0$ since fluctuations of density $\rho$, velocity $v$, and temperature $T$ are statistically independent \cite{Landau1980indp, Landau1980ipt}. Therefore we have
\begin{equation}\label{AP_ini delta condition}
\begin{split}
    \left<  {\rho}^2 \right>(0,x)&=\frac{m N_{eff}}{ \rho_0 L }\delta(x) \\
    \left<  {\rho v} \right>(0,x)&=0 \\
    \left<  {\rho T} \right>(0,x)&=0, \\
\end{split}
\end{equation}
in which $m$ is the mass of gas molecule, $N_{eff}$ is the effective number of molecules per particle used in the DSMC simulation taking its Monte Carlo fluctuation into account, $\rho_0$ is the equilibrium gas density, $L$ is the reference length scale used in the non-dimensionalization.

To solve the linear system \eqref{AP_Non-dim Macroscopic Equations 2}, we take the Fourier transform on the spacial coordinate and the one sided Fourier transformation on the temporal coordinate. With the help of the constitutive relation  \eqref{AP_The Linear Model on Derivatives: stress_and_flux, Fourier rescale} we obtain

\begin{equation}\label{AP_The DSMC Result:  with linear model, Fourier}
	\begin{split}
		i \omega \left<     { \rho}^2 \right>_{\omega,k}^{+} +  i k  \left<     { \rho}  {v} \right>_{\omega,k}^{+}  &= \frac{mN_{eff}}{2\pi \rho_0 L } \\
		i \omega \left<     { \rho}  {v} \right>_{\omega,k}^{+} + i k \left<     { \rho}  { T} \right>_{\omega,k}^{+} + i k  \left<     { \rho}^2 \right>_{\omega,k}^{+} &=  
		 -    k^2 A_k(\Kn)    \left<     { \rho}  {v} \right>_{\omega,k}^{+} \\
		\frac{3 }{2}i \omega \left<     { \rho}  { T} \right>_{\omega,k}^{+} + i k  \left<     { \rho}  {v} \right>_{\omega,k}^{+}  &= 
		-  k^2 B_k(\Kn) \left<     { \rho}  { T} \right>_{\omega,k}^{+} \\
	\end{split}
\end{equation}
in which we define $A_k(\Kn) = \mbox{Kn}M(k\Kn)$, $B_k(\Kn) = \frac{15}{4}\mbox{Kn}K(k\Kn)$, and $f^+_{\omega, k}$ for arbitrary function  $f(t,x)$ is obtained by taking the one sided Fourier transformation on time coordinate $t$ and the Fourier transform on space coordinate $x$. 

Finally, the solution of $\left<  \delta  {\rho}^2 \right>_{\omega,k}^{+}$ is a function of the Knudsen number $\Kn$ and the frequency $\omega$ only
\begin{widetext}
\begin{equation}\label{AP_The Linear Model on Derivatives: solution}
    	\left<  {\rho}^2 \right>_{\omega,k=2\pi}^{+}(\Kn)= 
	-\frac{i k N_{eff} m \left(-2 k^4 A_k(\text{Kn}) B_k(\text{Kn})-3 i k^2 \omega  A_k(\text{Kn})-2 i k^2 \omega  B_k(\text{Kn})-2 k^2+3 \omega ^2\right)}{(2 \pi)^2 \rho_0  \left(-2 k^4 \omega  A_k(\text{Kn}) B_k(\text{Kn})-3 i k^2 \omega ^2 A_k(\text{Kn})+2 i k^4 B_k(\text{Kn})-2 i k^2 \omega ^2 B_k(\text{Kn})-5 k^2 \omega +3 \omega ^3\right)},
\end{equation}
\end{widetext}
it does not depend on $k$ because that the non-dimensionalized wavenumber $k$ will be fixed to $2\pi$ by choosing the reference length scale $L=\frac{2\pi}{|k_0|}$ if we consider the Fourier mode at wavenumber $k_0$. The density fluctuation spectra $\left<\rho^2\right>(\omega, \Kn)$ is two times of the real part of \eqref{AP_The Linear Model on Derivatives: solution} according to \eqref{AP_Spectral Resolution of fluctuations: The FT, one sided 2}. This concludes the derivation of density fluctuation spectra for the constitutive relation \eqref{AP_The Linear Model on Derivatives: stress_and_flux, Fourier rescale}.

Next we determine the density fluctuation spectra for the NS equation and the Grad 13 moment method. The density fluctuation spectra computed from the NS equation could be obtained by simply replace $A_k, B_k$ in \eqref{AP_The Linear Model on Derivatives: solution} by
\begin{equation} \label{AP_NS spectra}	
A_k(\Kn) = \frac{4}{3}\Kn;\quad B_k(\Kn) = \frac{15}{4}\Kn
\end{equation}

As for the Grad 13 method, $\sigma$ and $q$ are unknown quantities determined by two addition equation for the stress and heat flux \cite[Eq 35]{Wu2020} as

\begin{equation}\label{AP_The DSMC Result: Grad 13}
	\begin{split}
		\Kn \partial_{ {t}}  {\sigma} +  \frac{4}{3}\partial_{ {x}}   {v}_x + \frac{8}{15}\partial_{ {x}}  {q} &= -\sigma\\
		\Kn\partial_{ {t}}   {q} + \frac{4}{15}\partial_{ {x}}  {\sigma}+ \frac{2}{3}\partial_{ {x}}    {T} &= -\frac{2}{3}q 
	\end{split}
\end{equation}
Note that the non-dimensinoalization used in \cite{Wu2020} differ with us for stress and heat flux, specifically we have $\sigma_{[26]} = \sigma \Kn $, $q_{[26]} = \frac{15}{4}q\Kn  $.

We apply the one sided Fourier transformation on time and Fourier transformation on space coordinates to the linear system again. The resulting governing equation for correlations of density between density, velocity, temperature, stress and heat fluxes for the Grad 13 are as follows

\begin{equation}\label{AP_The DSMC Result: Grad 13, Fourier}
	\begin{split}
		i \omega \left<    {  \rho}^2\right>_{\omega,k}^{+} + i k \left<    {  \rho} {v}_x\right>_{\omega,k}^{+}  &= \frac{mN_{eff}}{2\pi  \rho_0 L}\\
i \omega \left<    {  \rho} {v}_x\right>_{\omega,k}^{+} + i k\left<    {  \rho}   {  T}\right>_{\omega,k}^{+} + i k \left<    {  \rho}^2\right>_{\omega,k}^{+}  &= - i k \Kn \left<    {  \rho} {\sigma}\right>_{\omega,k}^{+}\\
\frac{3 }{2}i \omega \left<    {  \rho}   {  T}\right>_{\omega,k}^{+} + i k  \left<    {  \rho} {v}_x\right>_{\omega,k}^{+} + i k \left<    {  \rho} {q}\right>_{\omega,k}^{+} &= 0\\
i \omega \Kn \left<    {  \rho} {\sigma}\right>_{\omega,k}^{+} +  \frac{4}{3}i k  \left<    {  \rho} {v}_x\right>_{\omega,k}^{+} + \frac{8}{15}i k \left<    {  \rho} {q}\right>_{\omega,k}^{+} &= -\left<    {  \rho} {\sigma}\right>_{\omega,k}^{+}\\
i \omega \Kn  \left<    {  \rho} {q}\right>_{\omega,k}^{+} + \frac{4}{15}i k \left<    {  \rho} {\sigma}\right>_{\omega,k}^{+}+ \frac{2 }{3}i k \left<    {  \rho}   {  T}\right>_{\omega,k}^{+} &= -\frac{2}{3}\left<    {  \rho} {q}\right>_{\omega,k}^{+}\\			
\end{split}
\end{equation}
The spectra is hence calculated in the same way as NS equation and linear constitutional relation model. The result for the Grad 13 method is 
\begin{widetext}
\begin{equation}\label{AP_The DSMC Result: grad 13 solution}
	\left<  \delta {\rho}^2 \right>_{\omega,k=2\pi}^{+}(\Kn) = \frac{m N_{\text{eff}} k\left(-36 i k^4 \text{Kn}^2+189 i k^2 \text{Kn}^2 \omega ^2+165 k^2 \text{Kn} \omega -20 i k^2-45 i \text{Kn}^2 \omega ^4-75 \text{Kn} \omega ^3+30 i \omega ^2\right)}{(2 \pi)^2  \rho _0 \left(135 k^4 \text{Kn}^2 \omega -75 i k^4 \text{Kn}-234 k^2 \text{Kn}^2 \omega ^3+240 i k^2 \text{Kn} \omega ^2+50 k^2 \omega +45 \text{Kn}^2 \omega ^5-75 i \text{Kn} \omega ^4-30 \omega ^3\right)}
\end{equation}
\end{widetext}
Finally, we compute the velocity fluctuation spectra as the test data. Taking the correlation between velocity $v$ and the governing equation \eqref{AP_Non-dim Macroscopic Equations} gives
\begin{equation}\label{AP_Non-dim Macroscopic Equations 2 velo}
	\begin{split}
		\frac{\partial \left< {v \rho} \right>}{\partial  {t}} + \frac{\partial \left< { v^2} \right>}{\partial  {x}}     &= 0\\
		\frac{\partial  {\left< v^2 \right>}}{\partial  {t}} + \frac{\partial  \left<{v T} \right>}{\partial  {x}} + \frac{\partial  \left<{v \rho}\right>}{\partial  {x}} &=   -\mbox{Kn}  \frac{\partial   \sigma( \left<{ v^2}\right> )}{\partial  {x}}  \\
		\frac{3 }{2}\frac{\partial  \left<{v T}\right>}{\partial  {t}} + \frac{\partial  \left<{v \rho}\right>}{\partial  {x}} &= -  \frac{15 }{4 } \mbox{Kn}\frac{\partial   q(\left<{v T}\right>)}{\partial  {x}},
	\end{split}
\end{equation} 
The initial condition for this linear system is obtained following a similar argument with the density spectra case. At time $t=0$, the only non-vanishing correlation function is $ \left<  {v^2} \right>$, which is proportional to $\delta(x)$. The initial condition for $ \left<  {v}^2 \right>(0, x)$ we used here is deduced from \cite[Eq 88.5]{Landau1980indp} with non-dimensionalization and DSMC's Monte Carlo effects considered.
\begin{equation}\label{AP_ini delta condition v}
\begin{split}
    \left<  {v \rho} \right>(0,x)&=0 \\
    \left<  {v^2} \right>(0,x)&=\frac{m N_{eff}}{ \rho_0 L }\delta(x) \\
    \left<  {v T} \right>(0,x)&=0, \\
\end{split}
\end{equation}
%\begin{widetext}
%\begin{equation}\label{AP_The Linear Model on Derivatives:  velocity }
%	\begin{split}
%		i \omega \left<     { \rho}  {v}_x \right>_{\omega, k}^{+} +  ik  \left<    {v}_x^2 \right>_{\omega, k}^{+}  &=  0\\
%		i \omega \left<    {v}_x^2 \right>_{\omega, k}^{+} + ik \left<    {v}_x  {\delta T} \right>_{\omega, k}^{+} + ik  \left<    {\delta \rho}  {v}_x  \right>_{\omega, k}^{+} &=    - k^2 A(\textit{Kn})      \left<     {\delta \rho}  {v}_x \right>_{\omega, k}^{+} +\frac{N_{eff} m k}{(2\pi)^2 \delta \rho_0 }  \\
%		\frac{3 }{2}i \omega \left<    {v}_x  {\delta T} \right>_{\omega, k}^{+} + ik  \left<    {v}_x^2 \right>_{\omega, k}^{+}  &= - k^2 B(\textit{Kn})  \left<   {v}_x  {\delta T} \right>_{\omega, k}^{+}  \\
%	\end{split}
%\end{equation}
%\end{widetext}
\begin{widetext}
\begin{equation}\label{AP_The Linear Model on Derivatives: velocity result}
	\left<   {v}^2 \right>_{\omega,k=2\pi}^{+}(\Kn) =-\frac{i N_{eff} m \omega k  \left(3 \omega -2 i k^2 B_k(\text{Kn})\right)}{(2\pi)^2 \rho_0  (-2 k^4 \omega  A_k(\text{Kn}) B_k(\text{Kn})-3 i k^2 \omega ^2 A_k(\text{Kn})+2 i k^4 B_k(\text{Kn})-2 i k^2 \omega ^2 B_k(\text{Kn})-5 k^2 \omega +3 \omega ^3)}
\end{equation}
\end{widetext}
The velocity fluctuation spectra $\left<v^2\right>(\omega, \Kn)$ is two times of the real part of \eqref{AP_The Linear Model on Derivatives: velocity result} according to \eqref{AP_Spectral Resolution of fluctuations: The FT, one sided 2}.

\section{DSMC Calculation Details} \label{Ap: E}
We use the DSMC0F program by A.Bird \cite{Bird1994MolecularGDF} to simulate the fluctuation of 1D homogeneous gas. Its geometry is a one-dimensional gas of unit cross section between two plane specularly reflecting walls that are normal to the x-axis. The computation domain between these two plane has spatial span $4.8m$ in the x direction and is divided uniformly into $1281$ cells. Each cell contains $8$ subcells utilized in determining collision pairs in the DSMC computation.

The initial condition of our DSMC computation uses particle velocity sampled as in \cite{Bird1994MolecularGDFRD} from the Maxwell distribution at $T=300K$ and zero mean velocity. The particle position is uniformly distributed in each cell. More details about the properties of the gas are shown in Table \ref{AP_tab:table2} using SI units. 

The merit of the DSMC calculation is that no driven physical conditions are required for simulating fluctuations. It is because the DSMC method uses Monte Carlo samples to mimic the real gas molecules. Hence statistical quantities computed from such Monte Carlo samples naturally fluctuate in the same way as the real gas except for an enlarged fluctuation amplitude. Specifically, if one sample in the DSMC simulation represents $N_{eff}$ real gas molecules, the variances of fluctuations in statistical quantities computed from the DSMC simulation are $N_{eff}$ times larger than those of real gass. In our DSMC computation we have $128100$ simulation particle samples representing gases of number density $10^{20}m^{-3}$, therefore in our computation each sample particles representing $N_{eff}=3.75\times 10^{15}$ real gas molecules.

No global Knudsen number is defined for our DSMC simulation of homogeneous gas since there is no mean flow across the simulation domain. However, fluctuations in density, velocity, and temperature exist and propagate according to hydrodynamics with well-defined Knudsen numbers. Specifically, the Knudsen numbers are defined for various Fourier modes (Phonons) of the fluctuations according to their wavelength.

The molecular model is crutial in DSMC calculations. It describes how two molecule collide with each other and determines the viscosity of the gas. The molecular model gives the relation between two characteristic quantities of classical binary collision problem \cite{Bird1994Molecularcollide, griffiths_schroeter_2018collide,landau1982mechanics}: the impact parameter $b$ and scattering angle $\theta$. One of the typical molecular model used in DSMC is the variable hard/soft sphere model \cite{Bird1994MolecularMModel}

\begin{equation} \label{VHS model: The deflection angle}
	\theta = 2 \cos^{-1}((\frac{b}{d})^{\frac{1}{\alpha}})
\end{equation}
in which $d$ is the effective diameter of the gas molecules and $\alpha$ is a parameter mainly effecting the diffusion coefficient. The diffusion parameter describes mass diffusion between components of gas mixtures and is irrelevant in our single species case. Therefore we use the default value $\alpha=1$ in the DSMC0F program corresponding to the variable hard sphere model. The effective diameter $d$ varies with the relative velocity between colliding molecules according to \cite[Eq 4.63]{Bird1994MolecularEffd}
\begin{equation} \label{DSMC: the effective diameter, calculation, final}
	d = d_{ref}(\frac{(2k_B T_{ref}/(\frac{1}{2}m v_r^2))^{\omega-1/2}}{\Gamma(5/2-\omega)})^{1/2}
\end{equation}
in which $m$ is the mass of gas molecule, $v_r$ is the relative velocity between the two colliding molecules, $\Gamma$ represents the Gamma function, $T_{ref}$ is the reference temperature, $d_{ref}$ is the reference molecule diameter, and $\omega$ is a parameter determines how viscosity coefficient changes w.r.t temperature. In our computation we use the default value $m=5\times10^{-26} kg$, $T_{ref}=273K$, and $d_{ref}=3.5 \times 10^{-10}m$ in the DSMC0F program. Note that \eqref{DSMC: the effective diameter, calculation, final} differ from the equation in \cite{Bird1994MolecularEffd} since they use the reduced mass $m_r=\frac{1}{2}m$ instead of molecule mass $m$ in our case.

The parameter $\omega$ in \eqref{DSMC: the effective diameter, calculation, final} determines the power law between viscosity coefficient $\mu$ and the temperature $T$ in the form $\mu \propto T^\omega$ \cite[Eq 3.66]{Bird1994MolecularVisc}. However, viscosity coefficients appears only in the stress as a production with the velocity gradients. Such a change in viscosity is of second order $\delta v \delta T$ in the perturbation expansion hence is not important in our first-order linear case. As a result, we again use the default value $\omega=0.5$ in the DSMC0F program. 

We compute the viscosity coefficient and heat conduction coefficient of our DSMC simulated gas using the Chapman-Enskog theory \cite[Eq 3.73]{Bird1994MolecularVisc}
\begin{equation} \label{DSMC viscosity coef}
\begin{split}
    	\mu_0 &= \frac{5(\alpha+1)(\alpha+2)(\pi m k_B)^{1/2}(4k_B/m)^{\omega-1/2} T^\omega}{16\alpha\Gamma(\frac{9}{2}-\omega)\sigma_{T,ref}v_{r,ref}^{2\omega-1}}\\
	\kappa_0 &= \frac{15 k_B}{4 m}\mu_0
\end{split}
\end{equation}
in which the reference total cross section $\sigma_{T,ref} =\pi d_{ref}^2 $ and the reference velocity $v_{r,ref}  = \sqrt{\frac{4k_B T_{ref}}{m\Gamma(5/2-\omega)^{\frac{1}{\omega-1/2}}}}$. 

To ensure the resolution at relative large Knudsen numbers, our DSMC computation, use the cell width to be 5 times smaller than the mean free path of the gas, while the time step to be 10 times smaller than the mean free time of the gas.  We compute mean free path and mean free time from the collision rate per gas molecule according to \cite[Eq 4.64]{Bird1994MolecularMft}
\begin{equation} \label{DSMC: the collision frequency, general, deff, explicit, numerical}
	f = 4 n \pi^{1/2} d_{ref}^2  (\frac{ T}{T_{ref}})^{1/2-\omega}(\frac{k_B T}{m})^{1/2}
\end{equation}
in which $n$ is the number density. The mean free time of gas molecules is $t_m = \frac{1}{f}$, while the mean free path of gas molecules is $l_m = \sqrt{\frac{8 k_B T}{\pi m}} t_m \approx 1.28 l$, in which $l$ is the mean free path used in non-dimensionalization in Appendix \ref{Ap A}. 

There is no need to worry about the convergence in the mean flow since our computation simulates homogeneous gas using homogeneous initial conditions. However, the finite simulation domain in our DSMC computation may introduce deviation in the spectra from theoretical results in Appendix \ref{Ap: D}. To eliminate this finite domain effect, we use a domain length much (300 times) larger than the mean free path of the gas. Moreover, random perturbations that possibly appear in the initial condition are fully relaxed since we use the total simulation time of $30$ times more than the transverse time of sound speed over the computation domain.

A snap shot will be stored for every 5 time steps. Then the macroscopic quantities for each cell are calculated by averaging corresponding quantites of particles in each cell. The density fluctuation spectra used to train the neural network is ensemble average from 27 independent DSMC run, computed via discrete Fourier transformation using the equation \eqref{AP_Spectral Resolution of fluctuations: The spectra, meaning}.

\begin{table}
\caption{\label{AP_tab:table2}
The coefficients and configuration used in DSMC0F program in SI unit.}
\begin{ruledtabular}
\begin{tabular}{c|c|c|c}
Domain Length & 4.8 & Collision Model & VHS\footnotemark[1]\\
Power law\footnotemark[2] & 0.5 & Diameter\footnotemark[3] & $3.5\times10^{-10}$\\
Num of Cell & 1281 & Mean Free Path & $1.8\times10^{-2}$ \\
Simulation Particle & 128100 & Mean Free Time & $4\times10^{-5}$ \\
Density & $5\times10^{-6}$ & Temperature & 300 \\
Molecule Mass & $5\times10^{-26}$ & Sound Speed & 371.56 \\
Heat Conduction \footnotemark[5]& 0.0214 & Viscosity \footnotemark[6] &$2.07\times10^{-5}$ \\
Time step size & $4\times10^{-6}$ & Cell width & $3.7\times10^{-3}$\\
Subcell \footnotemark[4]& 8 & Simulation time& $4\times10^{-1}$ \\
\end{tabular}
\end{ruledtabular}
\footnotetext[1]{Variable hard sphere model}
\footnotetext[2]{The viscosity-temperature power law used in variable hard sphere model}
\footnotetext[3]{The reference molecule diameter}
\footnotetext[4]{The number of subcell per cell used in particle collision process}
\footnotetext[5]{The heat conduction coefficient}
\footnotetext[6]{The viscosity coefficient}
\end{table}

\section{Neural Network Training Details} \label{AP: F}
In this section, we adopt the dimensionalized quantities instead of non-dimensionalized version in Appendix \ref{Ap A} to make this section consistent with the DSMC computation which is computed in SI unit. In the dimensionalized notation, the density fluctuation spectra is of the form $\left<\delta \rho^2\right>(\omega,k)$, in which $\omega$ is the frequency and the wavenumber $k$. The wavenumber $k$ directly determines the Knudsen number of phonons (Fourier modes of $\rho$). Given $k$, the spectra $\left<\rho^2\right>_k(\omega)$ is a function of the angular frequency $\omega$.

The training data set consists of 40000 $(k,\omega, \left<\rho^2\right>)$ tuples draw. While the validation set consists 400 such tuples. We generate these tuples by draw $k$ uniformly from the interval $\left[0,\frac{\pi\rho_0}{2\mu}\sqrt{\frac{k_B T_0}{m}}\right]$ (corresponds to $\Kn \in [0,0.25]$). Then for a given $k$, we sample  $\omega$ from the range $[-3ck, 3ck]$ ($c$ is the speed of sound) with probability proportional to the value of $\left<\rho^2\right>_k(\omega)$ calculated using DSMC. The merit of such sampling strategy is it emphasis the peak region of the spectra. 

The common practice of choosing test set is to sample tuples of $(k,\omega, \left<\rho^2\right>)$ from the same distribution with the training set. However, such test set only test how good the neural network fit the density fluctuation spectra, not its ability to generalize to other physics scenarios. Instead, we use the model trained on density fluctuation spectra to predict the velocity fluctuation spectra $\left<v^2\right>(\omega, k)$ to test its generalization ability.

The function $M$ is modeled as a fully connected neural network without bias, as shown in the paper. The weights to be trained are $\mathbf{W}_{1,2}$. These weights are initialized by Pytorch's default uniform initialization. 

The loss is defined as the mean square difference between spectra $\left<\rho^2\right>_{DSMC}$ computed using DSMC and the spectra $\left<\rho^2\right>$ predicted by linear constitutiont relation model. The optimizer we use is the Adam optimizer with learning rate $\alpha = 0.005$, Beta parameter $\beta_1 = 0.9$ and $\beta_2 = 0.999$, and the parameter $\epsilon = 10^{-8}$. For each training epoch, the batch size for each step is 64. The training process stops if the loss obtained on validation set increases.

\bibliography{apssamp}% Produces the bibliography via BibTeX.

\providecommand{\noopsort}[1]{}\providecommand{\singleletter}[1]{#1}%
\begin{thebibliography}{61}%
\makeatletter
\providecommand \@ifxundefined [1]{%
 \@ifx{#1\undefined}
}%
\providecommand \@ifnum [1]{%
 \ifnum #1\expandafter \@firstoftwo
 \else \expandafter \@secondoftwo
 \fi
}%
\providecommand \@ifx [1]{%
 \ifx #1\expandafter \@firstoftwo
 \else \expandafter \@secondoftwo
 \fi
}%
\providecommand \natexlab [1]{#1}%
\providecommand \enquote  [1]{``#1''}%
\providecommand \bibnamefont  [1]{#1}%
\providecommand \bibfnamefont [1]{#1}%
\providecommand \citenamefont [1]{#1}%
\providecommand \href@noop [0]{\@secondoftwo}%
\providecommand \href [0]{\begingroup \@sanitize@url \@href}%
\providecommand \@href[1]{\@@startlink{#1}\@@href}%
\providecommand \@@href[1]{\endgroup#1\@@endlink}%
\providecommand \@sanitize@url [0]{\catcode `\\12\catcode `\$12\catcode
  `\&12\catcode `\#12\catcode `\^12\catcode `\_12\catcode `\%12\relax}%
\providecommand \@@startlink[1]{}%
\providecommand \@@endlink[0]{}%
\providecommand \url  [0]{\begingroup\@sanitize@url \@url }%
\providecommand \@url [1]{\endgroup\@href {#1}{\urlprefix }}%
\providecommand \urlprefix  [0]{URL }%
\providecommand \Eprint [0]{\href }%
\providecommand \doibase [0]{https://doi.org/}%
\providecommand \selectlanguage [0]{\@gobble}%
\providecommand \bibinfo  [0]{\@secondoftwo}%
\providecommand \bibfield  [0]{\@secondoftwo}%
\providecommand \translation [1]{[#1]}%
\providecommand \BibitemOpen [0]{}%
\providecommand \bibitemStop [0]{}%
\providecommand \bibitemNoStop [0]{.\EOS\space}%
\providecommand \EOS [0]{\spacefactor3000\relax}%
\providecommand \BibitemShut  [1]{\csname bibitem#1\endcsname}%
\let\auto@bib@innerbib\@empty
%</preamble>
\bibitem [{\citenamefont {Chen}\ and\ \citenamefont
  {Doolen}(1998)}]{Chen1998LATTICEBM}%
  \BibitemOpen
  \bibfield  {author} {\bibinfo {author} {\bibfnamefont {S.}~\bibnamefont
  {Chen}}\ and\ \bibinfo {author} {\bibfnamefont {G.~D.}\ \bibnamefont
  {Doolen}},\ }\bibfield  {title} {\bibinfo {title} {Lattice boltzmann method
  for fluid flows},\ }\href@noop {} {\bibfield  {journal} {\bibinfo  {journal}
  {Annual Review of Fluid Mechanics}\ }\textbf {\bibinfo {volume} {30}},\
  \bibinfo {pages} {329} (\bibinfo {year} {1998})}\BibitemShut {NoStop}%
\bibitem [{\citenamefont {Peter}\ and\ \citenamefont
  {Kremer}(2009)}]{Peter2009MultiscaleSO}%
  \BibitemOpen
  \bibfield  {author} {\bibinfo {author} {\bibfnamefont {C.}~\bibnamefont
  {Peter}}\ and\ \bibinfo {author} {\bibfnamefont {K.}~\bibnamefont {Kremer}},\
  }\bibfield  {title} {\bibinfo {title} {Multiscale simulation of soft matter
  systems – from the atomistic to the coarse-grained level and back},\
  }\href@noop {} {\bibfield  {journal} {\bibinfo  {journal} {Soft Matter}\
  }\textbf {\bibinfo {volume} {5}},\ \bibinfo {pages} {4357} (\bibinfo {year}
  {2009})}\BibitemShut {NoStop}%
\bibitem [{\citenamefont {Lin}\ and\ \citenamefont
  {Truhlar}(2006)}]{Lin2006QMMMWH}%
  \BibitemOpen
  \bibfield  {author} {\bibinfo {author} {\bibfnamefont {H.}~\bibnamefont
  {Lin}}\ and\ \bibinfo {author} {\bibfnamefont {D.~G.}\ \bibnamefont
  {Truhlar}},\ }\bibfield  {title} {\bibinfo {title} {Qm/mm: what have we
  learned, where are we, and where do we go from here?},\ }\href@noop {}
  {\bibfield  {journal} {\bibinfo  {journal} {Theoretical Chemistry Accounts}\
  }\textbf {\bibinfo {volume} {117}},\ \bibinfo {pages} {185} (\bibinfo {year}
  {2006})}\BibitemShut {NoStop}%
\bibitem [{\citenamefont {Kogan}(1969)}]{Kogan1969RarefiedGD}%
  \BibitemOpen
  \bibfield  {author} {\bibinfo {author} {\bibfnamefont {M.~N.}\ \bibnamefont
  {Kogan}},\ }\bibinfo {title} {Introduction},\ in\ \href
  {https://doi.org/10.1007/978-1-4899-6381-9_1} {\emph {\bibinfo {booktitle}
  {Rarefied Gas Dynamics}}}\ (\bibinfo  {publisher} {Springer US},\ \bibinfo
  {address} {Boston, MA},\ \bibinfo {year} {1969})\ pp.\ \bibinfo {pages}
  {1--27}\BibitemShut {NoStop}%
\bibitem [{\citenamefont {Bird}(1994{\natexlab{a}})}]{Bird1994}%
  \BibitemOpen
  \bibfield  {author} {\bibinfo {author} {\bibfnamefont {G.}~\bibnamefont
  {Bird}},\ }\href {https://books.google.com.hk/books?id=Bya5QgAACAAJ} {\emph
  {\bibinfo {title} {Molecular Gas Dynamics and the Direct Simulation of Gas
  Flows}}}\ (\bibinfo  {publisher} {Clarendon Press},\ \bibinfo {year}
  {1994})\BibitemShut {NoStop}%
\bibitem [{\citenamefont {Agarwal}\ \emph {et~al.}(2001)\citenamefont
  {Agarwal}, \citenamefont {Yun},\ and\ \citenamefont
  {Balakrishnan}}]{agarwal2001beyond}%
  \BibitemOpen
  \bibfield  {author} {\bibinfo {author} {\bibfnamefont {R.~K.}\ \bibnamefont
  {Agarwal}}, \bibinfo {author} {\bibfnamefont {K.-Y.}\ \bibnamefont {Yun}},\
  and\ \bibinfo {author} {\bibfnamefont {R.}~\bibnamefont {Balakrishnan}},\
  }\bibfield  {title} {\bibinfo {title} {Beyond navier--stokes: Burnett
  equations for flows in the continuum--transition regime},\ }\href@noop {}
  {\bibfield  {journal} {\bibinfo  {journal} {Physics of Fluids}\ }\textbf
  {\bibinfo {volume} {13}},\ \bibinfo {pages} {3061} (\bibinfo {year}
  {2001})}\BibitemShut {NoStop}%
\bibitem [{\citenamefont {Garc{\'i}a-Col{\'i}n}\ \emph
  {et~al.}(2008)\citenamefont {Garc{\'i}a-Col{\'i}n}, \citenamefont {Velasco},\
  and\ \citenamefont {Uribe}}]{GarcaColn2008BeyondTN}%
  \BibitemOpen
  \bibfield  {author} {\bibinfo {author} {\bibfnamefont {L.~S.}\ \bibnamefont
  {Garc{\'i}a-Col{\'i}n}}, \bibinfo {author} {\bibfnamefont {R.~M.}\
  \bibnamefont {Velasco}},\ and\ \bibinfo {author} {\bibfnamefont {F.~J.}\
  \bibnamefont {Uribe}},\ }\bibfield  {title} {\bibinfo {title} {Beyond the
  navier-stokes equations: Burnett hydrodynamics},\ }\href@noop {} {\bibfield
  {journal} {\bibinfo  {journal} {Physics Reports}\ }\textbf {\bibinfo {volume}
  {465}},\ \bibinfo {pages} {149} (\bibinfo {year} {2008})}\BibitemShut
  {NoStop}%
\bibitem [{\citenamefont {Hilbert}(1912)}]{Hilbert1912}%
  \BibitemOpen
  \bibfield  {author} {\bibinfo {author} {\bibfnamefont {D.}~\bibnamefont
  {Hilbert}},\ }\bibfield  {title} {\bibinfo {title} {{Begr{\"{u}}ndung der
  kinetischen Gastheorie}},\ }\bibfield  {journal} {\bibinfo  {journal}
  {Mathematische Annalen}\ }\href {https://doi.org/10.1007/BF01456676}
  {10.1007/BF01456676} (\bibinfo {year} {1912})\BibitemShut {NoStop}%
\bibitem [{Cha(1916)}]{Chapman1916}%
  \BibitemOpen
  \bibfield  {title} {\bibinfo {title} {{VI. On the law of distribution of
  molecular velocities, and on the theory of viscosity and thermal conduction,
  in a non-uniform simple monatomic gas}},\ }\bibfield  {journal} {\bibinfo
  {journal} {Philosophical Transactions of the Royal Society of London. Series
  A, Containing Papers of a Mathematical or Physical Character}\ }\href
  {https://doi.org/10.1098/rsta.1916.0006} {10.1098/rsta.1916.0006} (\bibinfo
  {year} {1916})\BibitemShut {NoStop}%
\bibitem [{\citenamefont {Enskog}(1917)}]{enskog1917kinetische}%
  \BibitemOpen
  \bibfield  {author} {\bibinfo {author} {\bibfnamefont {D.}~\bibnamefont
  {Enskog}},\ }\bibfield  {title} {\bibinfo {title} {Kinetische theorie der
  vorg{\"a}nge in m{\"a}ssig verd{\"u}nnten gasen. i. allgemeiner teil},\
  }\href@noop {} {\bibfield  {journal} {\bibinfo  {journal} {Uppsala: Almquist
  \& Wiksells Boktryckeri}\ } (\bibinfo {year} {1917})}\BibitemShut {NoStop}%
\bibitem [{\citenamefont {Bobylev}(1982)}]{bobylev1982chapman}%
  \BibitemOpen
  \bibfield  {author} {\bibinfo {author} {\bibfnamefont {A.}~\bibnamefont
  {Bobylev}},\ }\bibfield  {title} {\bibinfo {title} {The chapman-enskog and
  grad methods for solving the boltzmann equation},\ }in\ \href@noop {} {\emph
  {\bibinfo {booktitle} {Akademiia Nauk SSSR Doklady}}},\ Vol.\ \bibinfo
  {volume} {262}\ (\bibinfo {year} {1982})\ pp.\ \bibinfo {pages}
  {71--75}\BibitemShut {NoStop}%
\bibitem [{\citenamefont {McLennan}(1965)}]{McLennan1965ConvergenceOT}%
  \BibitemOpen
  \bibfield  {author} {\bibinfo {author} {\bibfnamefont {J.~A.}\ \bibnamefont
  {McLennan}},\ }\bibfield  {title} {\bibinfo {title} {Convergence of the
  chapman‐enskog expansion for the linearized boltzmann equation},\
  }\href@noop {} {\bibfield  {journal} {\bibinfo  {journal} {Physics of
  Fluids}\ }\textbf {\bibinfo {volume} {8}},\ \bibinfo {pages} {1580} (\bibinfo
  {year} {1965})}\BibitemShut {NoStop}%
\bibitem [{\citenamefont {Grad}(1949)}]{Grad1949}%
  \BibitemOpen
  \bibfield  {author} {\bibinfo {author} {\bibfnamefont {H.}~\bibnamefont
  {Grad}},\ }\bibfield  {title} {\bibinfo {title} {On the kinetic theory of
  rarefied gases},\ }\href
  {https://doi.org/https://doi.org/10.1002/cpa.3160020403} {\bibfield
  {journal} {\bibinfo  {journal} {Communications on Pure and Applied
  Mathematics}\ }\textbf {\bibinfo {volume} {2}},\ \bibinfo {pages} {331}
  (\bibinfo {year} {1949})}\BibitemShut {NoStop}%
\bibitem [{\citenamefont {Struchtrup}\ and\ \citenamefont
  {Taheri}(2011)}]{Struchtrup2011}%
  \BibitemOpen
  \bibfield  {author} {\bibinfo {author} {\bibfnamefont {H.}~\bibnamefont
  {Struchtrup}}\ and\ \bibinfo {author} {\bibfnamefont {P.}~\bibnamefont
  {Taheri}},\ }\bibfield  {title} {\bibinfo {title} {{Macroscopic transport
  models for rarefied gas flows: A brief review}},\ }\href
  {https://doi.org/10.1093/imamat/hxr004} {\bibfield  {journal} {\bibinfo
  {journal} {IMA Journal of Applied Mathematics (Institute of Mathematics and
  Its Applications)}\ }\textbf {\bibinfo {volume} {76}},\ \bibinfo {pages}
  {672} (\bibinfo {year} {2011})}\BibitemShut {NoStop}%
\bibitem [{\citenamefont {Weiss}(1995)}]{weiss1995continuous}%
  \BibitemOpen
  \bibfield  {author} {\bibinfo {author} {\bibfnamefont {W.}~\bibnamefont
  {Weiss}},\ }\bibfield  {title} {\bibinfo {title} {Continuous shock structure
  in extended thermodynamics},\ }\href@noop {} {\bibfield  {journal} {\bibinfo
  {journal} {Physical Review E}\ }\textbf {\bibinfo {volume} {52}},\ \bibinfo
  {pages} {R5760} (\bibinfo {year} {1995})}\BibitemShut {NoStop}%
\bibitem [{\citenamefont {Torrilhon}(2009)}]{Torrilhon2009HyperbolicME}%
  \BibitemOpen
  \bibfield  {author} {\bibinfo {author} {\bibfnamefont {M.}~\bibnamefont
  {Torrilhon}},\ }\bibfield  {title} {\bibinfo {title} {Hyperbolic moment
  equations in kinetic gas theory based on multi-variate
  pearson-iv-distributions},\ }\href@noop {} {\bibfield  {journal} {\bibinfo
  {journal} {Communications in Computational Physics}\ }\textbf {\bibinfo
  {volume} {7}},\ \bibinfo {pages} {639} (\bibinfo {year} {2009})}\BibitemShut
  {NoStop}%
\bibitem [{\citenamefont {Hana}\ \emph {et~al.}(2019)\citenamefont {Hana},
  \citenamefont {Ma}, \citenamefont {Ma},\ and\ \citenamefont
  {Weinan}}]{Hana2019}%
  \BibitemOpen
  \bibfield  {author} {\bibinfo {author} {\bibfnamefont {J.}~\bibnamefont
  {Hana}}, \bibinfo {author} {\bibfnamefont {C.}~\bibnamefont {Ma}}, \bibinfo
  {author} {\bibfnamefont {Z.}~\bibnamefont {Ma}},\ and\ \bibinfo {author}
  {\bibfnamefont {E.}~\bibnamefont {Weinan}},\ }\bibfield  {title} {\bibinfo
  {title} {{Uniformly accurate machine learning-based hydrodynamic models for
  kinetic equations}},\ }\bibfield  {journal} {\bibinfo  {journal} {Proceedings
  of the National Academy of Sciences of the United States of America}\ }\href
  {https://doi.org/10.1073/pnas.1909854116} {10.1073/pnas.1909854116} (\bibinfo
  {year} {2019})\BibitemShut {NoStop}%
\bibitem [{\citenamefont {Zhang}\ and\ \citenamefont {Ma}(2020)}]{Zhang2020}%
  \BibitemOpen
  \bibfield  {author} {\bibinfo {author} {\bibfnamefont {J.}~\bibnamefont
  {Zhang}}\ and\ \bibinfo {author} {\bibfnamefont {W.}~\bibnamefont {Ma}},\
  }\bibfield  {title} {\bibinfo {title} {{Data-driven discovery of governing
  equations for fluid dynamics based on molecular simulation}},\ }\bibfield
  {journal} {\bibinfo  {journal} {Journal of Fluid Mechanics}\ }\href
  {https://doi.org/10.1017/jfm.2020.184} {10.1017/jfm.2020.184} (\bibinfo
  {year} {2020})\BibitemShut {NoStop}%
\bibitem [{\citenamefont {Raissi}\ \emph {et~al.}(2017)\citenamefont {Raissi},
  \citenamefont {Perdikaris},\ and\ \citenamefont {Karniadakis}}]{Raissi2017}%
  \BibitemOpen
  \bibfield  {author} {\bibinfo {author} {\bibfnamefont {M.}~\bibnamefont
  {Raissi}}, \bibinfo {author} {\bibfnamefont {P.}~\bibnamefont {Perdikaris}},\
  and\ \bibinfo {author} {\bibfnamefont {G.~E.}\ \bibnamefont {Karniadakis}},\
  }\bibfield  {title} {\bibinfo {title} {{Machine learning of linear
  differential equations using Gaussian processes}},\ }\bibfield  {journal}
  {\bibinfo  {journal} {Journal of Computational Physics}\ }\href
  {https://doi.org/10.1016/j.jcp.2017.07.050} {10.1016/j.jcp.2017.07.050}
  (\bibinfo {year} {2017}),\ \Eprint {https://arxiv.org/abs/1701.02440}
  {arXiv:1701.02440} \BibitemShut {NoStop}%
\bibitem [{\citenamefont {Rudy}\ \emph {et~al.}(2017)\citenamefont {Rudy},
  \citenamefont {Brunton}, \citenamefont {Proctor},\ and\ \citenamefont
  {Kutz}}]{Rudy2017}%
  \BibitemOpen
  \bibfield  {author} {\bibinfo {author} {\bibfnamefont {S.~H.}\ \bibnamefont
  {Rudy}}, \bibinfo {author} {\bibfnamefont {S.~L.}\ \bibnamefont {Brunton}},
  \bibinfo {author} {\bibfnamefont {J.~L.}\ \bibnamefont {Proctor}},\ and\
  \bibinfo {author} {\bibfnamefont {J.~N.}\ \bibnamefont {Kutz}},\ }\bibfield
  {title} {\bibinfo {title} {{Data-driven discovery of partial differential
  equations}},\ }\bibfield  {journal} {\bibinfo  {journal} {Science Advances}\
  }\href {https://doi.org/10.1126/sciadv.1602614} {10.1126/sciadv.1602614}
  (\bibinfo {year} {2017}),\ \Eprint {https://arxiv.org/abs/1609.06401}
  {arXiv:1609.06401} \BibitemShut {NoStop}%
\bibitem [{\citenamefont {Schaeffer}(2017)}]{Schaeffer2017}%
  \BibitemOpen
  \bibfield  {author} {\bibinfo {author} {\bibfnamefont {H.}~\bibnamefont
  {Schaeffer}},\ }\bibfield  {title} {\bibinfo {title} {{Learning partial
  differential equations via data discovery and sparse optimization}},\
  }\bibfield  {journal} {\bibinfo  {journal} {Proceedings of the Royal Society
  A: Mathematical, Physical and Engineering Sciences}\ }\href
  {https://doi.org/10.1098/rspa.2016.0446} {10.1098/rspa.2016.0446} (\bibinfo
  {year} {2017})\BibitemShut {NoStop}%
\bibitem [{\citenamefont {Raissi}\ \emph {et~al.}(2019)\citenamefont {Raissi},
  \citenamefont {Perdikaris},\ and\ \citenamefont {Karniadakis}}]{Raissi2019}%
  \BibitemOpen
  \bibfield  {author} {\bibinfo {author} {\bibfnamefont {M.}~\bibnamefont
  {Raissi}}, \bibinfo {author} {\bibfnamefont {P.}~\bibnamefont {Perdikaris}},\
  and\ \bibinfo {author} {\bibfnamefont {G.~E.}\ \bibnamefont {Karniadakis}},\
  }\bibfield  {title} {\bibinfo {title} {{Physics-informed neural networks: A
  deep learning framework for solving forward and inverse problems involving
  nonlinear partial differential equations}},\ }\bibfield  {journal} {\bibinfo
  {journal} {Journal of Computational Physics}\ }\href
  {https://doi.org/10.1016/j.jcp.2018.10.045} {10.1016/j.jcp.2018.10.045}
  (\bibinfo {year} {2019})\BibitemShut {NoStop}%
\bibitem [{\citenamefont {Long}\ \emph {et~al.}(2018)\citenamefont {Long},
  \citenamefont {Lu}, \citenamefont {Ma},\ and\ \citenamefont
  {Dong}}]{Long2018}%
  \BibitemOpen
  \bibfield  {author} {\bibinfo {author} {\bibfnamefont {Z.}~\bibnamefont
  {Long}}, \bibinfo {author} {\bibfnamefont {Y.}~\bibnamefont {Lu}}, \bibinfo
  {author} {\bibfnamefont {X.}~\bibnamefont {Ma}},\ and\ \bibinfo {author}
  {\bibfnamefont {B.}~\bibnamefont {Dong}},\ }\bibfield  {title} {\bibinfo
  {title} {{PDE-Net: Learning PDEs from data}},\ }in\ \href@noop {} {\emph
  {\bibinfo {booktitle} {6th International Conference on Learning
  Representations, ICLR 2018 - Workshop Track Proceedings}}}\ (\bibinfo {year}
  {2018})\BibitemShut {NoStop}%
\bibitem [{\citenamefont {Long}\ \emph {et~al.}(2019)\citenamefont {Long},
  \citenamefont {Lu},\ and\ \citenamefont {Dong}}]{Long2019}%
  \BibitemOpen
  \bibfield  {author} {\bibinfo {author} {\bibfnamefont {Z.}~\bibnamefont
  {Long}}, \bibinfo {author} {\bibfnamefont {Y.}~\bibnamefont {Lu}},\ and\
  \bibinfo {author} {\bibfnamefont {B.}~\bibnamefont {Dong}},\ }\bibfield
  {title} {\bibinfo {title} {{PDE-Net 2.0: Learning PDEs from data with a
  numeric-symbolic hybrid deep network}},\ }\bibfield  {journal} {\bibinfo
  {journal} {Journal of Computational Physics}\ }\textbf {\bibinfo {volume}
  {399}},\ \href {https://doi.org/10.1016/j.jcp.2019.108925}
  {10.1016/j.jcp.2019.108925} (\bibinfo {year} {2019}),\ \Eprint
  {https://arxiv.org/abs/1812.04426} {arXiv:1812.04426} \BibitemShut {NoStop}%
\bibitem [{\citenamefont {LeCun}\ \emph {et~al.}(1995)\citenamefont {LeCun},
  \citenamefont {Bengio} \emph {et~al.}}]{lecun1995convolutional}%
  \BibitemOpen
  \bibfield  {author} {\bibinfo {author} {\bibfnamefont {Y.}~\bibnamefont
  {LeCun}}, \bibinfo {author} {\bibfnamefont {Y.}~\bibnamefont {Bengio}}, \emph
  {et~al.},\ }\bibfield  {title} {\bibinfo {title} {Convolutional networks for
  images, speech, and time series},\ }\href@noop {} {\bibfield  {journal}
  {\bibinfo  {journal} {The handbook of brain theory and neural networks}\
  }\textbf {\bibinfo {volume} {3361}},\ \bibinfo {pages} {1995} (\bibinfo
  {year} {1995})}\BibitemShut {NoStop}%
\bibitem [{\citenamefont {{Dal Santo}}\ \emph {et~al.}(2020)\citenamefont {{Dal
  Santo}}, \citenamefont {Deparis},\ and\ \citenamefont
  {Pegolotti}}]{DalSanto2020}%
  \BibitemOpen
  \bibfield  {author} {\bibinfo {author} {\bibfnamefont {N.}~\bibnamefont {{Dal
  Santo}}}, \bibinfo {author} {\bibfnamefont {S.}~\bibnamefont {Deparis}},\
  and\ \bibinfo {author} {\bibfnamefont {L.}~\bibnamefont {Pegolotti}},\
  }\bibfield  {title} {\bibinfo {title} {{Data driven approximation of
  parametrized PDEs by reduced basis and neural networks}},\ }\href
  {https://doi.org/10.1016/j.jcp.2020.109550} {\bibfield  {journal} {\bibinfo
  {journal} {Journal of Computational Physics}\ }\textbf {\bibinfo {volume}
  {416}},\ \bibinfo {pages} {109550} (\bibinfo {year} {2020})},\ \Eprint
  {https://arxiv.org/abs/1904.01514} {arXiv:1904.01514} \BibitemShut {NoStop}%
\bibitem [{\citenamefont {Baydin}\ \emph {et~al.}(2018)\citenamefont {Baydin},
  \citenamefont {Pearlmutter}, \citenamefont {Radul},\ and\ \citenamefont
  {Siskind}}]{baydin2018automatic}%
  \BibitemOpen
  \bibfield  {author} {\bibinfo {author} {\bibfnamefont {A.~G.}\ \bibnamefont
  {Baydin}}, \bibinfo {author} {\bibfnamefont {B.~A.}\ \bibnamefont
  {Pearlmutter}}, \bibinfo {author} {\bibfnamefont {A.~A.}\ \bibnamefont
  {Radul}},\ and\ \bibinfo {author} {\bibfnamefont {J.~M.}\ \bibnamefont
  {Siskind}},\ }\bibfield  {title} {\bibinfo {title} {Automatic differentiation
  in machine learning: a survey},\ }\href@noop {} {\bibfield  {journal}
  {\bibinfo  {journal} {Journal of machine learning research}\ }\textbf
  {\bibinfo {volume} {18}} (\bibinfo {year} {2018})}\BibitemShut {NoStop}%
\bibitem [{\citenamefont {Yip}\ and\ \citenamefont {Nelkin}(1964)}]{Yip1964}%
  \BibitemOpen
  \bibfield  {author} {\bibinfo {author} {\bibfnamefont {S.}~\bibnamefont
  {Yip}}\ and\ \bibinfo {author} {\bibfnamefont {M.}~\bibnamefont {Nelkin}},\
  }\bibfield  {title} {\bibinfo {title} {{Application of a kinetic model to
  time-dependent density correlations in fluids}},\ }\bibfield  {journal}
  {\bibinfo  {journal} {Physical Review}\ }\textbf {\bibinfo {volume} {135}},\
  \href {https://doi.org/10.1103/PhysRev.135.A1241} {10.1103/PhysRev.135.A1241}
  (\bibinfo {year} {1964})\BibitemShut {NoStop}%
\bibitem [{\citenamefont {Fiocco}\ and\ \citenamefont
  {DeWolf}(1968)}]{Fiocco1968}%
  \BibitemOpen
  \bibfield  {author} {\bibinfo {author} {\bibfnamefont {G.}~\bibnamefont
  {Fiocco}}\ and\ \bibinfo {author} {\bibfnamefont {J.}~\bibnamefont
  {DeWolf}},\ }\bibfield  {title} {\bibinfo {title} {Frequency spectrum of
  laser echoes from atmospheric constituents and determination of the aerosol
  content of air},\ }\href@noop {} {\bibfield  {journal} {\bibinfo  {journal}
  {Journal of Atmospheric Sciences}\ }\textbf {\bibinfo {volume} {25}},\
  \bibinfo {pages} {488} (\bibinfo {year} {1968})}\BibitemShut {NoStop}%
\bibitem [{\citenamefont {Wu}\ and\ \citenamefont {Gu}(2020)}]{Wu2020}%
  \BibitemOpen
  \bibfield  {author} {\bibinfo {author} {\bibfnamefont {L.}~\bibnamefont
  {Wu}}\ and\ \bibinfo {author} {\bibfnamefont {X.-J.}\ \bibnamefont {Gu}},\
  }\bibfield  {title} {\bibinfo {title} {{On the accuracy of macroscopic
  equations for linearized rarefied gas flows}},\ }\bibfield  {journal}
  {\bibinfo  {journal} {Advances in Aerodynamics}\ }\textbf {\bibinfo {volume}
  {2}},\ \href {https://doi.org/10.1186/s42774-019-0025-4}
  {10.1186/s42774-019-0025-4} (\bibinfo {year} {2020})\BibitemShut {NoStop}%
\bibitem [{\citenamefont {Chapman}\ and\ \citenamefont
  {Cowling}(1990)}]{chapman1990mathematical}%
  \BibitemOpen
  \bibfield  {author} {\bibinfo {author} {\bibfnamefont {S.}~\bibnamefont
  {Chapman}}\ and\ \bibinfo {author} {\bibfnamefont {T.~G.}\ \bibnamefont
  {Cowling}},\ }\href@noop {} {\emph {\bibinfo {title} {The mathematical theory
  of non-uniform gases: an account of the kinetic theory of viscosity, thermal
  conduction and diffusion in gases}}}\ (\bibinfo  {publisher} {Cambridge
  university press},\ \bibinfo {year} {1990})\BibitemShut {NoStop}%
\bibitem [{\citenamefont {Lifshitz}\ and\ \citenamefont
  {Pitaevskii}(2013{\natexlab{a}})}]{Landau1980hf}%
  \BibitemOpen
  \bibfield  {author} {\bibinfo {author} {\bibfnamefont {E.~M.}\ \bibnamefont
  {Lifshitz}}\ and\ \bibinfo {author} {\bibfnamefont {L.~P.}\ \bibnamefont
  {Pitaevskii}},\ }\href@noop {} {\emph {\bibinfo {title} {Statistical physics:
  theory of the condensed state}}},\ Vol.~\bibinfo {volume} {9}\ (\bibinfo
  {publisher} {Elsevier},\ \bibinfo {year} {2013})\ Chap.~\bibinfo {chapter}
  {88}, p.\ \bibinfo {pages} {372}\BibitemShut {NoStop}%
\bibitem [{\citenamefont {Landau}\ and\ \citenamefont
  {Lifshitz}(2013)}]{Landau1980ipt}%
  \BibitemOpen
  \bibfield  {author} {\bibinfo {author} {\bibfnamefont {L.~D.}\ \bibnamefont
  {Landau}}\ and\ \bibinfo {author} {\bibfnamefont {E.~M.}\ \bibnamefont
  {Lifshitz}},\ }\href@noop {} {\emph {\bibinfo {title} {Course of theoretical
  physics}}},\ Vol.~\bibinfo {volume} {5}\ (\bibinfo  {publisher} {Elsevier},\
  \bibinfo {year} {2013})\ Chap.\ \bibinfo {chapter} {112}, p.\ \bibinfo
  {pages} {343}\BibitemShut {NoStop}%
\bibitem [{\citenamefont {Michalis}\ \emph {et~al.}(2010)\citenamefont
  {Michalis}, \citenamefont {Kalarakis}, \citenamefont {Skouras},\ and\
  \citenamefont {Burganos}}]{Michalis2010}%
  \BibitemOpen
  \bibfield  {author} {\bibinfo {author} {\bibfnamefont {V.}~\bibnamefont
  {Michalis}}, \bibinfo {author} {\bibfnamefont {A.}~\bibnamefont {Kalarakis}},
  \bibinfo {author} {\bibfnamefont {E.}~\bibnamefont {Skouras}},\ and\ \bibinfo
  {author} {\bibfnamefont {V.}~\bibnamefont {Burganos}},\ }\bibfield  {title}
  {\bibinfo {title} {Rarefaction effects on gas viscosity in the knudsen
  transition regime},\ }\href {https://doi.org/10.1007/s10404-010-0606-3}
  {\bibfield  {journal} {\bibinfo  {journal} {Microfluidics and Nanofluidics}\
  }\textbf {\bibinfo {volume} {9}},\ \bibinfo {pages} {847} (\bibinfo {year}
  {2010})}\BibitemShut {NoStop}%
\bibitem [{\citenamefont {Ghaem-Maghami}\ and\ \citenamefont
  {May}(1980)}]{GhaemMaghami1980RayleighBrillouinSO}%
  \BibitemOpen
  \bibfield  {author} {\bibinfo {author} {\bibfnamefont {V.}~\bibnamefont
  {Ghaem-Maghami}}\ and\ \bibinfo {author} {\bibfnamefont {A.~D.}\ \bibnamefont
  {May}},\ }\bibfield  {title} {\bibinfo {title} {Rayleigh-brillouin spectrum
  of compressed he, ne, and ar. ii. the hydrodynamic region},\ }\href@noop {}
  {\bibfield  {journal} {\bibinfo  {journal} {Physical Review A}\ }\textbf
  {\bibinfo {volume} {22}},\ \bibinfo {pages} {698} (\bibinfo {year}
  {1980})}\BibitemShut {NoStop}%
\bibitem [{\citenamefont {Goodfellow}\ \emph {et~al.}(2015)\citenamefont
  {Goodfellow}, \citenamefont {Bengio},\ and\ \citenamefont
  {Courville}}]{Goodfellow2015DeepL}%
  \BibitemOpen
  \bibfield  {author} {\bibinfo {author} {\bibfnamefont {I.~J.}\ \bibnamefont
  {Goodfellow}}, \bibinfo {author} {\bibfnamefont {Y.}~\bibnamefont {Bengio}},\
  and\ \bibinfo {author} {\bibfnamefont {A.~C.}\ \bibnamefont {Courville}},\
  }\bibfield  {title} {\bibinfo {title} {Deep learning},\ }\href@noop {}
  {\bibfield  {journal} {\bibinfo  {journal} {Nature}\ }\textbf {\bibinfo
  {volume} {521}},\ \bibinfo {pages} {436} (\bibinfo {year}
  {2015})}\BibitemShut {NoStop}%
\bibitem [{\citenamefont {Pecora}(1964)}]{Pecora1964}%
  \BibitemOpen
  \bibfield  {author} {\bibinfo {author} {\bibfnamefont {R.}~\bibnamefont
  {Pecora}},\ }\bibfield  {title} {\bibinfo {title} {{Doppler shifts in light
  scattering from pure liquids and polymer solutions}},\ }\href
  {https://doi.org/10.1063/1.1725368} {\bibfield  {journal} {\bibinfo
  {journal} {The Journal of Chemical Physics}\ }\textbf {\bibinfo {volume}
  {40}},\ \bibinfo {pages} {1604} (\bibinfo {year} {1964})}\BibitemShut
  {NoStop}%
\bibitem [{\citenamefont {Jackson}(1998{\natexlab{a}})}]{jackson1999classical}%
  \BibitemOpen
  \bibfield  {author} {\bibinfo {author} {\bibfnamefont {J.~D.}\ \bibnamefont
  {Jackson}},\ }\href@noop {} {\emph {\bibinfo {title} {Classical
  Electrodynamics, 3rd ed.}}}\ (\bibinfo  {publisher} {Wiley},\ \bibinfo {year}
  {1998})\ Chap.~\bibinfo {chapter} {10}\BibitemShut {NoStop}%
\bibitem [{\citenamefont {Landau}\ \emph {et~al.}(2013)\citenamefont {Landau},
  \citenamefont {Bell}, \citenamefont {Kearsley}, \citenamefont {Pitaevskii},
  \citenamefont {Lifshitz},\ and\ \citenamefont
  {Sykes}}]{landau2013electrodynamics}%
  \BibitemOpen
  \bibfield  {author} {\bibinfo {author} {\bibfnamefont {L.~D.}\ \bibnamefont
  {Landau}}, \bibinfo {author} {\bibfnamefont {J.}~\bibnamefont {Bell}},
  \bibinfo {author} {\bibfnamefont {M.}~\bibnamefont {Kearsley}}, \bibinfo
  {author} {\bibfnamefont {L.}~\bibnamefont {Pitaevskii}}, \bibinfo {author}
  {\bibfnamefont {E.}~\bibnamefont {Lifshitz}},\ and\ \bibinfo {author}
  {\bibfnamefont {J.}~\bibnamefont {Sykes}},\ }\href@noop {} {\emph {\bibinfo
  {title} {Electrodynamics of continuous media}}},\ Vol.~\bibinfo {volume} {8}\
  (\bibinfo  {publisher} {elsevier},\ \bibinfo {year} {2013})\ Chap.\ \bibinfo
  {chapter} {117}\BibitemShut {NoStop}%
\bibitem [{\citenamefont {Kingma}\ and\ \citenamefont {Ba}(2015)}]{Kingma2015}%
  \BibitemOpen
  \bibfield  {author} {\bibinfo {author} {\bibfnamefont {D.~P.}\ \bibnamefont
  {Kingma}}\ and\ \bibinfo {author} {\bibfnamefont {J.~L.}\ \bibnamefont
  {Ba}},\ }\bibfield  {title} {\bibinfo {title} {{Adam: A method for stochastic
  optimization}},\ }in\ \href@noop {} {\emph {\bibinfo {booktitle} {3rd
  International Conference on Learning Representations, ICLR 2015 - Conference
  Track Proceedings}}}\ (\bibinfo {year} {2015})\ \Eprint
  {https://arxiv.org/abs/1412.6980} {arXiv:1412.6980} \BibitemShut {NoStop}%
\bibitem [{\citenamefont {Roohi}\ and\ \citenamefont
  {Darbandi}(2009)}]{Roohi2009ExtendingTN}%
  \BibitemOpen
  \bibfield  {author} {\bibinfo {author} {\bibfnamefont {E.}~\bibnamefont
  {Roohi}}\ and\ \bibinfo {author} {\bibfnamefont {M.}~\bibnamefont
  {Darbandi}},\ }\bibfield  {title} {\bibinfo {title} {Extending the
  navier–stokes solutions to transition regime in two-dimensional micro- and
  nanochannel flows using information preservation scheme},\ }\href@noop {}
  {\bibfield  {journal} {\bibinfo  {journal} {Physics of Fluids}\ }\textbf
  {\bibinfo {volume} {21}},\ \bibinfo {pages} {082001} (\bibinfo {year}
  {2009})}\BibitemShut {NoStop}%
\bibitem [{\citenamefont {Karniadakis}\ \emph {et~al.}(2001)\citenamefont
  {Karniadakis}, \citenamefont {Beskok}, \citenamefont {Aluru},\ and\
  \citenamefont {Ho}}]{Karniadakis2001MicroflowsAN}%
  \BibitemOpen
  \bibfield  {author} {\bibinfo {author} {\bibfnamefont {G.~E.}\ \bibnamefont
  {Karniadakis}}, \bibinfo {author} {\bibfnamefont {A.}~\bibnamefont {Beskok}},
  \bibinfo {author} {\bibfnamefont {N.~R.}\ \bibnamefont {Aluru}},\ and\
  \bibinfo {author} {\bibfnamefont {C.-M.}\ \bibnamefont {Ho}},\ }\bibfield
  {title} {\bibinfo {title} {Microflows and nanoflows: Fundamentals and
  simulation}\ }(\bibinfo {year} {2001})\BibitemShut {NoStop}%
\bibitem [{\citenamefont {Grad}(1952)}]{Grad1952ThePO}%
  \BibitemOpen
  \bibfield  {author} {\bibinfo {author} {\bibfnamefont {H.}~\bibnamefont
  {Grad}},\ }\bibfield  {title} {\bibinfo {title} {The profile of a steady
  plane shock wave},\ }\href@noop {} {\bibfield  {journal} {\bibinfo  {journal}
  {Communications on Pure and Applied Mathematics}\ }\textbf {\bibinfo {volume}
  {5}},\ \bibinfo {pages} {257} (\bibinfo {year} {1952})}\BibitemShut {NoStop}%
\bibitem [{\citenamefont
  {Cercignani}(1988{\natexlab{a}})}]{Cercignani1988TheBE52}%
  \BibitemOpen
  \bibfield  {author} {\bibinfo {author} {\bibfnamefont {C.}~\bibnamefont
  {Cercignani}},\ }\bibinfo {title} {Small and large mean free paths},\ in\
  \href {https://doi.org/10.1007/978-1-4612-1039-9_5} {\emph {\bibinfo
  {booktitle} {The Boltzmann Equation and Its Applications}}}\ (\bibinfo
  {publisher} {Springer New York},\ \bibinfo {address} {New York, NY},\
  \bibinfo {year} {1988})\ p.\ \bibinfo {pages} {238}\BibitemShut {NoStop}%
\bibitem [{\citenamefont
  {Cercignani}(1988{\natexlab{b}})}]{Cercignani1988TheBE55}%
  \BibitemOpen
  \bibfield  {author} {\bibinfo {author} {\bibfnamefont {C.}~\bibnamefont
  {Cercignani}},\ }\bibinfo {title} {Small and large mean free paths},\ in\
  \href {https://doi.org/10.1007/978-1-4612-1039-9_5} {\emph {\bibinfo
  {booktitle} {The Boltzmann Equation and Its Applications}}}\ (\bibinfo
  {publisher} {Springer New York},\ \bibinfo {address} {New York, NY},\
  \bibinfo {year} {1988})\ p.\ \bibinfo {pages} {248}\BibitemShut {NoStop}%
\bibitem [{\citenamefont {Kardar}(2007)}]{Kardar2007StatisticalPOC37}%
  \BibitemOpen
  \bibfield  {author} {\bibinfo {author} {\bibfnamefont {M.}~\bibnamefont
  {Kardar}},\ }\bibfield  {title} {\bibinfo {title} {Statistical physics of
  particles}\ }(\bibinfo {year} {2007})\ Chap.~\bibinfo {chapter} {3}, pp.\
  \bibinfo {pages} {78--81}\BibitemShut {NoStop}%
\bibitem [{\citenamefont {Riley}\ \emph
  {et~al.}(2006{\natexlab{a}})\citenamefont {Riley}, \citenamefont {Hobson},\
  and\ \citenamefont {Bence}}]{riley_hobson_bence_2006}%
  \BibitemOpen
  \bibfield  {author} {\bibinfo {author} {\bibfnamefont {K.~F.}\ \bibnamefont
  {Riley}}, \bibinfo {author} {\bibfnamefont {M.~P.}\ \bibnamefont {Hobson}},\
  and\ \bibinfo {author} {\bibfnamefont {S.~J.}\ \bibnamefont {Bence}},\
  }\bibinfo {title} {Integral transforms},\ in\ \href
  {https://doi.org/10.1017/CBO9780511810763.016} {\emph {\bibinfo {booktitle}
  {Mathematical Methods for Physics and Engineering: A Comprehensive Guide}}}\
  (\bibinfo  {publisher} {Cambridge University Press},\ \bibinfo {year}
  {2006})\ Chap.~\bibinfo {chapter} {13}, p.\ \bibinfo {pages} {435},\ \bibinfo
  {edition} {3rd}\ ed.\BibitemShut {Stop}%
\bibitem [{\citenamefont {Lifshitz}\ and\ \citenamefont
  {Pitaevskii}(2013{\natexlab{b}})}]{Landau1980indp}%
  \BibitemOpen
  \bibfield  {author} {\bibinfo {author} {\bibfnamefont {E.~M.}\ \bibnamefont
  {Lifshitz}}\ and\ \bibinfo {author} {\bibfnamefont {L.~P.}\ \bibnamefont
  {Pitaevskii}},\ }\href@noop {} {\emph {\bibinfo {title} {Statistical physics:
  theory of the condensed state}}},\ Vol.~\bibinfo {volume} {9}\ (\bibinfo
  {publisher} {Elsevier},\ \bibinfo {year} {2013})\ Chap.~\bibinfo {chapter}
  {88}, p.\ \bibinfo {pages} {370}\BibitemShut {NoStop}%
\bibitem [{\citenamefont
  {Jackson}(1998{\natexlab{b}})}]{Jackson1999ClassicalE3}%
  \BibitemOpen
  \bibfield  {author} {\bibinfo {author} {\bibfnamefont {J.~D.}\ \bibnamefont
  {Jackson}},\ }\href@noop {} {\emph {\bibinfo {title} {Classical
  Electrodynamics, 3rd ed.}}}\ (\bibinfo  {publisher} {Wiley},\ \bibinfo {year}
  {1998})\ Chap.\ \bibinfo {chapter} {10.2}, p.\ \bibinfo {pages}
  {463}\BibitemShut {NoStop}%
\bibitem [{\citenamefont {Griffiths}\ and\ \citenamefont
  {Schroeter}(2018{\natexlab{a}})}]{griffiths_schroeter_2018}%
  \BibitemOpen
  \bibfield  {author} {\bibinfo {author} {\bibfnamefont {D.~J.}\ \bibnamefont
  {Griffiths}}\ and\ \bibinfo {author} {\bibfnamefont {D.~F.}\ \bibnamefont
  {Schroeter}},\ }\href {https://doi.org/10.1017/9781316995433} {\emph
  {\bibinfo {title} {Introduction to Quantum Mechanics}}},\ \bibinfo {edition}
  {3rd}\ ed.\ (\bibinfo  {publisher} {Cambridge University Press},\ \bibinfo
  {year} {2018})\ Chap.\ \bibinfo {chapter} {10.4}\BibitemShut {NoStop}%
\bibitem [{\citenamefont {Riley}\ \emph
  {et~al.}(2006{\natexlab{b}})\citenamefont {Riley}, \citenamefont {Hobson},\
  and\ \citenamefont {Bence}}]{riley_hobson_bence_2006_autoc}%
  \BibitemOpen
  \bibfield  {author} {\bibinfo {author} {\bibfnamefont {K.~F.}\ \bibnamefont
  {Riley}}, \bibinfo {author} {\bibfnamefont {M.~P.}\ \bibnamefont {Hobson}},\
  and\ \bibinfo {author} {\bibfnamefont {S.~J.}\ \bibnamefont {Bence}},\
  }\bibinfo {title} {Integral transforms},\ in\ \href
  {https://doi.org/10.1017/CBO9780511810763.016} {\emph {\bibinfo {booktitle}
  {Mathematical Methods for Physics and Engineering: A Comprehensive Guide}}}\
  (\bibinfo  {publisher} {Cambridge University Press},\ \bibinfo {year}
  {2006})\ Chap.~\bibinfo {chapter} {13}, p.\ \bibinfo {pages} {450},\ \bibinfo
  {edition} {3rd}\ ed.\BibitemShut {Stop}%
\bibitem [{\citenamefont {Lifshitz}\ and\ \citenamefont
  {Pitaevskii}(2013{\natexlab{c}})}]{Landau1980hfcal}%
  \BibitemOpen
  \bibfield  {author} {\bibinfo {author} {\bibfnamefont {E.~M.}\ \bibnamefont
  {Lifshitz}}\ and\ \bibinfo {author} {\bibfnamefont {L.~P.}\ \bibnamefont
  {Pitaevskii}},\ }\href@noop {} {\emph {\bibinfo {title} {Statistical physics:
  theory of the condensed state}}},\ Vol.~\bibinfo {volume} {9}\ (\bibinfo
  {publisher} {Elsevier},\ \bibinfo {year} {2013})\ Chap.~\bibinfo {chapter}
  {89}, pp.\ \bibinfo {pages} {373--377}\BibitemShut {NoStop}%
\bibitem [{\citenamefont {Bird}(1994{\natexlab{b}})}]{Bird1994MolecularGDF}%
  \BibitemOpen
  \bibfield  {author} {\bibinfo {author} {\bibfnamefont {G.}~\bibnamefont
  {Bird}},\ }\href {https://books.google.com.hk/books?id=Bya5QgAACAAJ} {\emph
  {\bibinfo {title} {Molecular Gas Dynamics and the Direct Simulation of Gas
  Flows}}}\ (\bibinfo  {publisher} {Clarendon Press},\ \bibinfo {year} {1994})\
  Chap.\ \bibinfo {chapter} {11.6}\BibitemShut {NoStop}%
\bibitem [{\citenamefont {Bird}(1994{\natexlab{c}})}]{Bird1994MolecularGDFRD}%
  \BibitemOpen
  \bibfield  {author} {\bibinfo {author} {\bibfnamefont {G.}~\bibnamefont
  {Bird}},\ }\href {https://books.google.com.hk/books?id=Bya5QgAACAAJ} {\emph
  {\bibinfo {title} {Molecular Gas Dynamics and the Direct Simulation of Gas
  Flows}}}\ (\bibinfo  {publisher} {Clarendon Press},\ \bibinfo {year} {1994})\
  Chap.\ \bibinfo {chapter} {Appendix C}, p.\ \bibinfo {pages}
  {426}\BibitemShut {NoStop}%
\bibitem [{\citenamefont
  {Bird}(1994{\natexlab{d}})}]{Bird1994Molecularcollide}%
  \BibitemOpen
  \bibfield  {author} {\bibinfo {author} {\bibfnamefont {G.}~\bibnamefont
  {Bird}},\ }\href {https://books.google.com.hk/books?id=Bya5QgAACAAJ} {\emph
  {\bibinfo {title} {Molecular Gas Dynamics and the Direct Simulation of Gas
  Flows}}}\ (\bibinfo  {publisher} {Clarendon Press},\ \bibinfo {year} {1994})\
  Chap.\ \bibinfo {chapter} {2.2}, p.~\bibinfo {pages} {33}\BibitemShut
  {NoStop}%
\bibitem [{\citenamefont {Griffiths}\ and\ \citenamefont
  {Schroeter}(2018{\natexlab{b}})}]{griffiths_schroeter_2018collide}%
  \BibitemOpen
  \bibfield  {author} {\bibinfo {author} {\bibfnamefont {D.~J.}\ \bibnamefont
  {Griffiths}}\ and\ \bibinfo {author} {\bibfnamefont {D.~F.}\ \bibnamefont
  {Schroeter}},\ }\href {https://doi.org/10.1017/9781316995433} {\emph
  {\bibinfo {title} {Introduction to Quantum Mechanics}}},\ \bibinfo {edition}
  {3rd}\ ed.\ (\bibinfo  {publisher} {Cambridge University Press},\ \bibinfo
  {year} {2018})\ Chap.\ \bibinfo {chapter} {10.1.1}, p.\ \bibinfo {pages}
  {376}\BibitemShut {NoStop}%
\bibitem [{\citenamefont {Landau}\ and\ \citenamefont
  {Lifshitz}(1982)}]{landau1982mechanics}%
  \BibitemOpen
  \bibfield  {author} {\bibinfo {author} {\bibfnamefont {L.}~\bibnamefont
  {Landau}}\ and\ \bibinfo {author} {\bibfnamefont {E.}~\bibnamefont
  {Lifshitz}},\ }\href {https://books.google.com.hk/books?id=bE-9tUH2J2wC}
  {\emph {\bibinfo {title} {Mechanics: Volume 1}}},\ \bibinfo {number} {v. 1}\
  (\bibinfo  {publisher} {Elsevier Science},\ \bibinfo {year} {1982})\
  Chap.~\bibinfo {chapter} {18}, p.~\bibinfo {pages} {48}\BibitemShut {NoStop}%
\bibitem [{\citenamefont {Bird}(1994{\natexlab{e}})}]{Bird1994MolecularMModel}%
  \BibitemOpen
  \bibfield  {author} {\bibinfo {author} {\bibfnamefont {G.}~\bibnamefont
  {Bird}},\ }\href {https://books.google.com.hk/books?id=Bya5QgAACAAJ} {\emph
  {\bibinfo {title} {Molecular Gas Dynamics and the Direct Simulation of Gas
  Flows}}}\ (\bibinfo  {publisher} {Clarendon Press},\ \bibinfo {year} {1994})\
  Chap.\ \bibinfo {chapter} {2.6-2.7}\BibitemShut {NoStop}%
\bibitem [{\citenamefont {Bird}(1994{\natexlab{f}})}]{Bird1994MolecularEffd}%
  \BibitemOpen
  \bibfield  {author} {\bibinfo {author} {\bibfnamefont {G.}~\bibnamefont
  {Bird}},\ }\href {https://books.google.com.hk/books?id=Bya5QgAACAAJ} {\emph
  {\bibinfo {title} {Molecular Gas Dynamics and the Direct Simulation of Gas
  Flows}}}\ (\bibinfo  {publisher} {Clarendon Press},\ \bibinfo {year} {1994})\
  Chap.\ \bibinfo {chapter} {4.3}, p.~\bibinfo {pages} {93}\BibitemShut
  {NoStop}%
\bibitem [{\citenamefont {Bird}(1994{\natexlab{g}})}]{Bird1994MolecularVisc}%
  \BibitemOpen
  \bibfield  {author} {\bibinfo {author} {\bibfnamefont {G.}~\bibnamefont
  {Bird}},\ }\href {https://books.google.com.hk/books?id=Bya5QgAACAAJ} {\emph
  {\bibinfo {title} {Molecular Gas Dynamics and the Direct Simulation of Gas
  Flows}}}\ (\bibinfo  {publisher} {Clarendon Press},\ \bibinfo {year} {1994})\
  Chap.\ \bibinfo {chapter} {3.5}, pp.\ \bibinfo {pages} {68--70}\BibitemShut
  {NoStop}%
\bibitem [{\citenamefont {Bird}(1994{\natexlab{h}})}]{Bird1994MolecularMft}%
  \BibitemOpen
  \bibfield  {author} {\bibinfo {author} {\bibfnamefont {G.}~\bibnamefont
  {Bird}},\ }\href {https://books.google.com.hk/books?id=Bya5QgAACAAJ} {\emph
  {\bibinfo {title} {Molecular Gas Dynamics and the Direct Simulation of Gas
  Flows}}}\ (\bibinfo  {publisher} {Clarendon Press},\ \bibinfo {year} {1994})\
  Chap.\ \bibinfo {chapter} {4.3}, p.~\bibinfo {pages} {93}\BibitemShut
  {NoStop}%
\end{thebibliography}%

\end{document}